\documentclass[usenatbib,usegraphicx,fleqn]{ctextemp_mn2e}
\usepackage{amsmath,amssymb,bm,color,multirow,multicol}
\usepackage{charter}   
\pdfoutput=1

\newcommand{\e}[1]{\times 10^{#1}}

\usepackage{ulem}

\newcommand{\dd}[2][]{\frac{d #1}{d #2}}                  

\title[Boundary between Stable and Unstable]{Boundary between Stable and Unstable Regimes of Accretion.
 Ordered and Chaotic Unstable Regimes}
\author[A. A. Blinova, M. M. Romanova, R. V. E. Lovelace]{A. A.
Blinova\thanks{E-mail: alisablinova@gmail.com}, M. M.
Romanova\thanks{E-mail: romanova@astro.cornell.edu},
R.V.E. Lovelace\thanks{E-mail: lovelace@astro.cornell.edu} \\
Dept. of Astronomy, Cornell University, Ithaca, NY 14853 }

\begin{document}
\maketitle

\begin{abstract}

We present a new study of the Rayleigh-Taylor unstable regime of
accretion onto rotating magnetized stars in a set of high grid
resolution  three-dimensional (3D) magnetohydrodynamic (MHD)
simulations performed in low-viscosity discs. We find that the
boundary between the stable and unstable regimes is determined
almost entirely by the fastness parameter
$\omega_s=\Omega_\star/\Omega_K(r_m)$, where $\Omega_\star$ is the
angular velocity of the star and $\Omega_K(r_m)$ is the angular
velocity of the Keplerian disc at the disc-magnetosphere boundary
$r=r_m$. We found that accretion is unstable if $\omega_s\lesssim
0.6$. Accretion through instabilities is present in stars with
different magnetospheric sizes. However, only in stars with
relatively small magnetospheres, $r_m/R_\star\lesssim 7$, do the
unstable tongues produce chaotic hot spots on the stellar surface
and irregular light-curves. At even smaller values of the fastness
parameter, $\omega_s\lesssim 0.45$, multiple irregular tongues
merge, forming one or two \textit{ordered unstable} tongues that
rotate with the angular frequency of the inner disc. This
transition occurs in stars with even smaller magnetospheres,
$r_m/{R_\star}\lesssim 4.2$. Most of our simulations were
performed at a small tilt of the dipole magnetosphere,
$\Theta=5^\circ$, and a small viscosity parameter $\alpha=0.02$.
Test simulations at higher  $\alpha$ values show that many more
cases become unstable, and the light-curves become even more
irregular. Test simulations at larger tilts of the dipole $\Theta$
show that instability is present, however, accretion in two funnel
streams dominates if $\Theta\gtrsim 15^\circ$. The results of
these simulations can be applied to accreting magnetized stars
with relatively small magnetospheres: Classical T Tauri stars,
accreting millisecond X-ray pulsars, and cataclysmics variables.

\end{abstract}



\begin{figure*}
\centering
\includegraphics[width=8cm,clip]{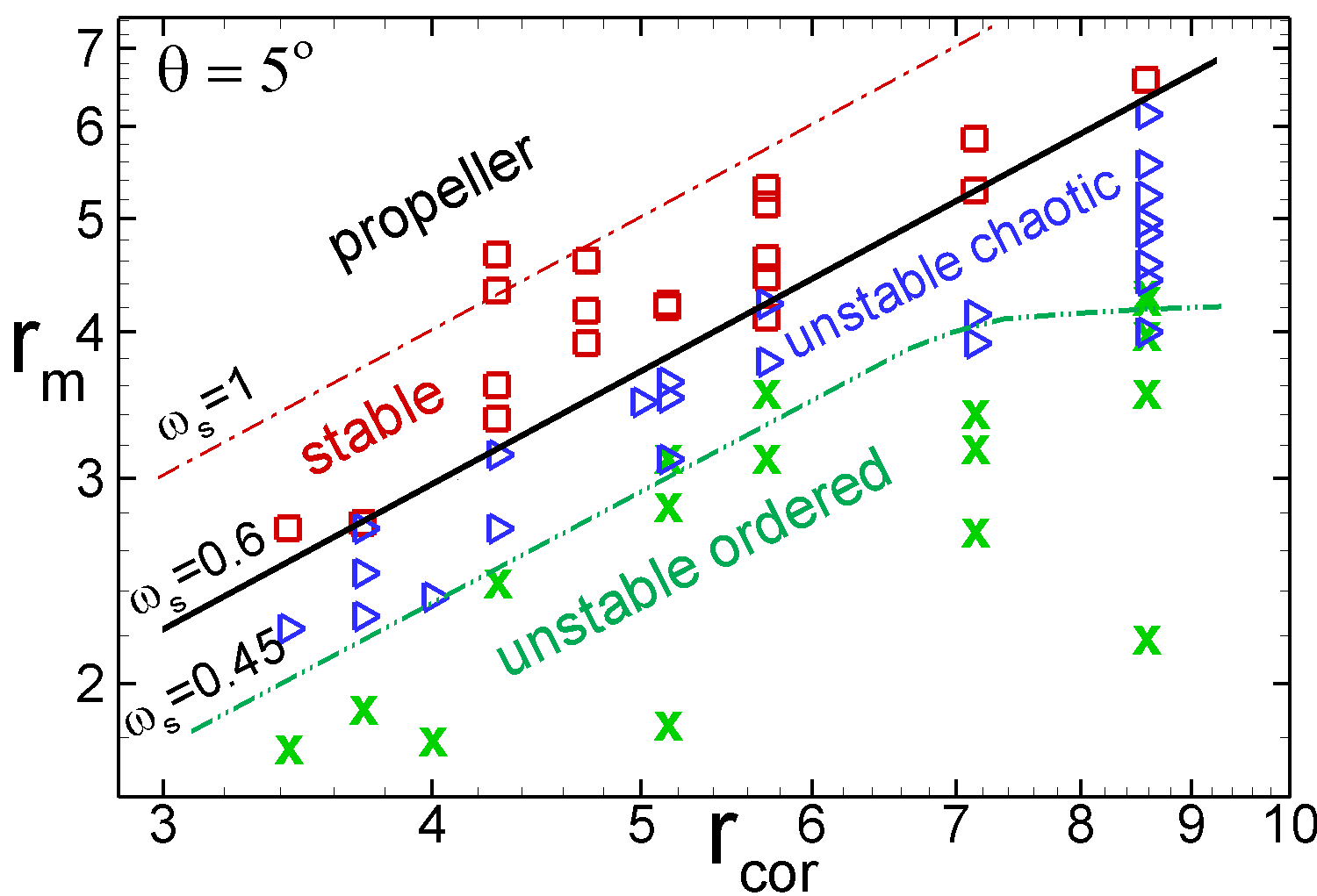}
\includegraphics[width=8cm,clip]{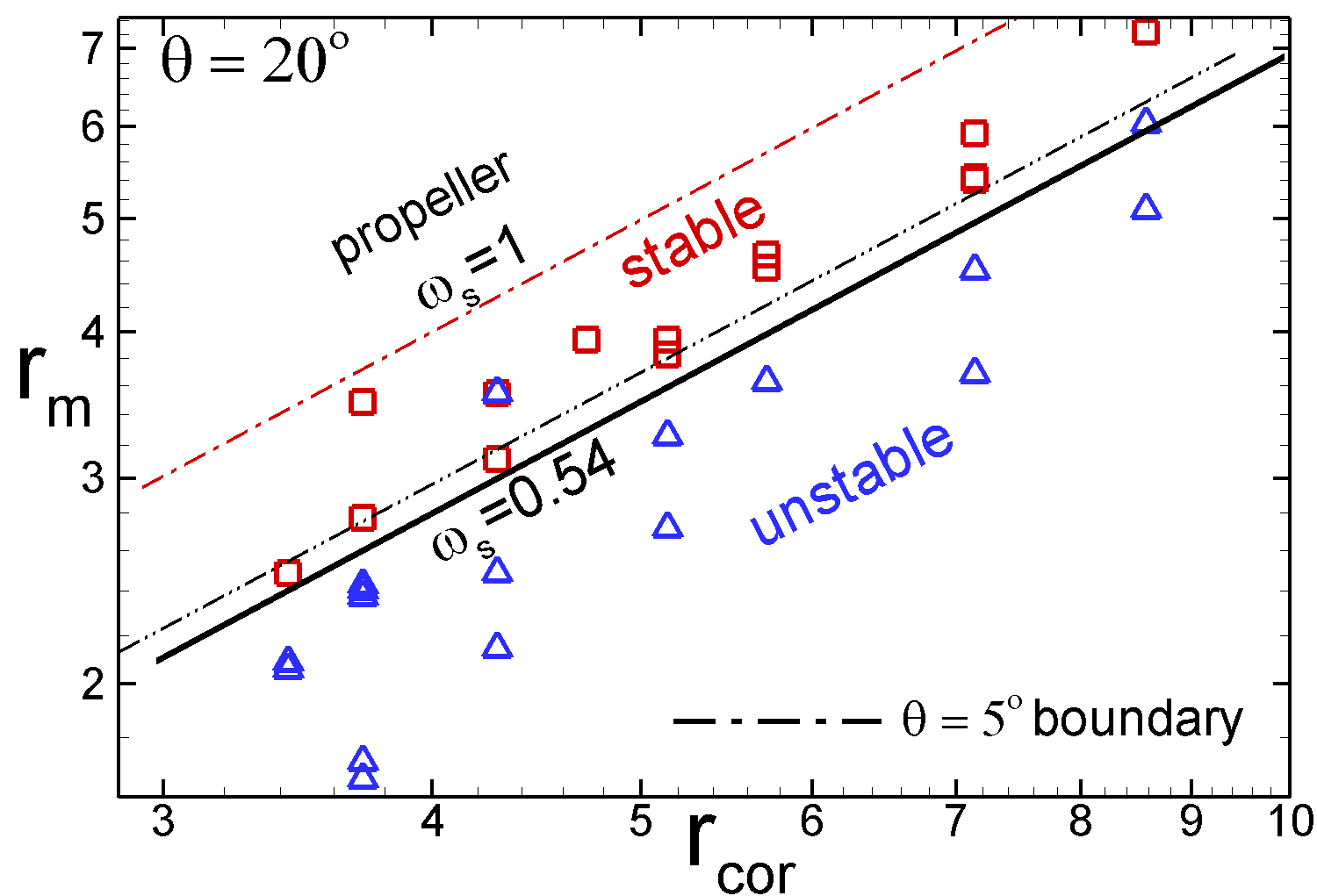}
\caption{The boundary between stable and unstable regimes of
accretion in the parameter space of $r_m$ and $r_{\rm cor}$ for
two different misalignment angles of the dipole: $\theta=5^\circ$
and $\theta=20^\circ$. Here, the units of length are given in
stellar radii for user convenience. \textit{Left panel}: boundary
line for $\Theta=5^\circ$. Stable, chaotic unstable, and ordered
unstable accretion cases are represented by red squares, blue
triangles, and green x's, respectively. \textit{Right panel}: the
boundary between stable and unstable regimes for
$\Theta=20^\circ$. The dash-dot-dot line shows the boundary for
$\theta=5^\circ$.} \label{boundary-t5-t20}
\end{figure*}

\section{Introduction}

Magnetospheric accretion occurs in different types of stars,
including Classical T Tauri stars (CTTSs) (e.g.,
\citealt{BouvierEtAl2007}), magnetized cataclysmic variables (CVs)
(e.g., \citealt{Warner1995,Hellier2001}), and accreting
millisecond X-ray pulsars (AMXPs)(e.g., \citealt{Vanderklis2006}).
The dynamically-important magnetic field stops the accretion disc
at some distance from the star, $r_m$ (called the magnetospheric
radius), and subsequently, the magnetic field governs the flow of
matter onto the star (e.g.,
\citealt{PringleRees1972,GhoshLamb1978,Campbell1992,LovelaceEtAl1995}).

The disc-magnetosphere boundary is prone to the Kelvin-Helmholtz
and magnetic Rayleigh-Taylor (RT) instabilities (e.g.,
\citealt{Chandrasekhar1961}, \citealt{AronsLea1976}). It was
suggested in earlier studies that the instabilities at the
disc-magnetosphere boundary may only lead to the mixing of plasma
with the field in the external layers of the magnetosphere
\citep{AronsLea1976}. However, our global 3D MHD simulations show
that the disc matter can deeply penetrate the magnetosphere in
tongues which produce chaotic hot spots on the stellar surface and
irregular light-curves. Simulations show that matter may accrete
in either the stable or the unstable regime. In the stable regime,
matter is lifted above the magnetosphere and accretes in two
ordered funnel streams, forming two ordered hot spots on the
surface of the star
\citep{RomanovaEtAl2003,RomanovaEtAl2004,KulkarniRomanova2005}. In
the unstable regime, matter penetrates through the magnetosphere
in several equatorial ``tongues" due to the magnetic
Rayleigh-Taylor instability
 and forms several chaotic spots on the surface of the
star
\citep{KulkarniRomanova2008,RomanovaEtAl2008,BachettiEtAl2010,KurosawaRomanova2013}.
The light-curves associated with the hot spots are expected to be
periodic in the stable regime, and irregular in the unstable
regime, with a typical time-scale of a few peaks per rotational
period of the inner disc.

Simulations performed locally in the 3D simulation box also show
that an ordered magnetic field is not an obstacle for the magnetic
Rayleigh-Taylor instability (e.g.,
\citealt{RastatterSchindler1999,StoneGardiner2007a,StoneGardiner2007b}).
These simulations and earlier simulations by
\citet{WangRobertson1984,WangRobertson1985} show that the
Rayleigh-Taylor instability leads to the formation of small-scale
waves and filaments that merge to form much larger filaments,
which then deeply penetrate the magnetically-dominated,
low-density regions.

The possible importance of the model for understanding CTTSs as
well as other magnetized stars (such as AMXPs and IPs) pushed us
to reconsider the unstable regime with new, more advanced
simulations at a higher grid resolution. Test simulations showed
that accretion through instabilities appeared in many more cases,
including those which were stable in simulations with the coarser
grid. For example, in the former simulations, accretion was
usually stable in the cases of a small $\alpha-$parameter of
viscosity \citep{ShakuraSunyaev1973}. However, in the new
simulations, many of these cases become unstable. We concluded
that the coarse grid was a factor that suppressed or weakened
instability in many of our earlier simulations, so we reconsidered
the problem using a higher grid resolution. We also tested the
convergence of the code at several grids with increasing
resolutions.

One of the important questions is, what determines the boundary
between stable and unstable regimes of accretion? In prior work,
we derived an empirical boundary that was based on the accretion
rate (which we varied using the $\alpha-$parameter of viscosity,
e.g. \citealt{RomanovaEtAl2008}). This boundary was useful,
although it has not been completely understood. This prior work
led to the important conclusion that models with higher values of
$\alpha$ are more unstable. One of the goals of our current
research is to derive a boundary between stable and unstable
regimes at a low viscosity in the disk.

Another important phenomenon that has been systematically observed
in the new simulations is the frequent presence of what we call
the \textit{ordered unstable regime}, where the unstable tongues
merge to form one or two ``ordered" tongues that rotate with the
frequency of the inner disc. The accretion in ordered unstable
regime has been observed in earlier simulations (performed at a
lower grid resolution), but only in the cases of very small
magnetospheres and high $\alpha-$parameters
\citep{RomanovaKulkarni2009}. In recent simulations at a higher
grid resolution, we observed that this regime is present at a
wider range of parameter values, including larger-sized
magnetospheres and lower viscosities in the disc.

In this paper, we focus on two main issues:  (1) searching a new
boundary between the stable and unstable regimes; ~ (2)
investigating the ordered unstable regime and finding a boundary
between the ordered and chaotic unstable regimes.

In Sec. 2 we discuss the theoretical background of the problem. In
Sec. 3 we describe the numerical model used in our simulations. In
Sec. 4 we search for the boundary between stable and unstable
regimes of accretion. In Sec. 5 we describe the ordered and
chaotic regimes of unstable accretion, and derive the boundary
between them. In Sec. 6 we analyze instability and investigate the
dependence of instability on the corotation radius and the
$\alpha-$parameter of viscosity. In Sec. 7 we investigate the grid
convergence of the model. In Sec. 8 we discuss the properties of
accretion through instabilities in the cases of larger tilts of
the dipole, $\Theta$. In Sec. 9 we compare relativistic and
non-relativistic cases. In Sec. 10 we discuss the applications to
different magnetized stars. We conclude in Sec. 11.

\begin{table*}
\centering
\begin{tabular}{llllllll}

\\ Model & $\mu$ & $r_{\rm cor}$ & $\alpha$ & $\Theta$ & $P_\star$  & $P_{\rm inst}$ & Comments\\
\hline
$\mu0.3c1.8\Theta5\alpha0.02$  & 0.3 & 1.8 & 0.02 & 5 & 2.4  & 1.5 & intermediate (unstable chaotic/stable)\\
$\mu0.3c2.5\Theta5\alpha0.02$  & 0.3 & 2.5 & 0.02 & 5 & 3.9 &  1.9 & unstable ordered\\
$\mu0.3c5\Theta10\alpha0.02$  & 0.3 & 5 & 0.02 & 10 & 11.2 &  2.2 & unstable ordered/chaotic\\
\hline
$\mu0.5c1.5\Theta5\alpha0.02$   & 0.5 & 1.5 &  0.02 & 5& 1.8  & 0.9 & unstable chaotic\\
$\mu0.5c2\Theta5\alpha0.02$ & 0.5 & 2 &  0.02 & 5 & 2.8  & 2.4 & unstable chaotic\\
$\mu0.5c3\Theta5\alpha0.02$ & 0.5 & 3 & 0.02 & 5 & 5.2 & 2.3, 2.7 & unstable ordered/chaotic \\
$\mu0.5c5\Theta5\alpha0.02$  & 0.5 & 5 &  0.02 & 5 & 11.2 & 3 & unstable ordered\\
$\mu0.5c3\Theta5\alpha0.1$  & 0.5 & 3 &  0.1 & 5 & 5.2 & 1.6 & unstable chaotic\\
$\mu0.5c5\Theta5\alpha0.1$  & 0.5 & 5 & 0.1 & 5 & 11.2  &  2.6 & unstable ordered\\
$\mu0.5c3\Theta20\alpha0.02$  & 0.5 & 3 &  0.02 & 20 & 5.2 & 1.9, 2.7 & unstable chaotic/ordered \\
\hline
$\mu1c1.8\Theta5\alpha0.02$  & 1 & 1.8 & 0.02 & 5 & 2.4 & 0.85 & intermediate (unstable chaotic/stable)\\
$\mu1c2\Theta5\alpha0.02$  & 1 & 2 & 0.02 & 5 & 2.8  & 0.4-1.6 & intermediate (unstable chaotic/stable)\\
$\mu1c2.5\Theta5\alpha0.02$ & 1& 2.5 & 0.02 & 5 & 3.9 & 3 & unstable chaotic\\
$\mu1c3\Theta5\alpha0.02$  & 1 & 3 & 0.02 & 5 & 5.2 & 1.9, 3.8 & unstable ordered\\
$\mu1c5\Theta5\alpha0.02$  & 1 & 5 & 0.02 & 5 & 11.2  & 1.8, 7.4 & unstable ordered\\
$\mu1c1.5\Theta5\alpha0.1$  & 1 & 1.5 & 0.1 & 5 & 1.8  & 0.9 & unstable chaotic\\
$\mu1c3\Theta5\alpha0.1$  & 1 & 3 & 0.1 & 5 & 5.2 & 2.7 & unstable chaotic \\
\hline
$\mu2c2\Theta5\alpha0.02$  & 2 & 2 & 0.02 & 5 & 5.2 & 2.8 & stable\\
$\mu2c3\Theta5\alpha0.02$  & 2 & 3 & 0.02 & 5 & 5.2 & 2.8 & unstable chaotic\\
\hline
\end{tabular}
\caption{Sample (representative) models for different values of
parameters $\mu$, $r_{\rm cor}$, $\alpha$,  and $\Theta$. The
period of the star $P_\star$ and the periods of instabilities
$P_{\rm inst}$ obtained in Fourier spectral analysis are shown.}
\label{tab:models}
\end{table*}

\section{Theoretical background}

\subsection{Disk-magnetosphere boundary.}

 The magnetic field of the star truncates
the accretion disk at a radius  $r_m$ where the magnetic stress in
the magnetosphere matches the matter stress in the disk (e.g.,
\citealt{PringleRees1972,LambEtAl1973}):
\begin{equation}
p + \rho v^2 = B^2/8\pi,~~~~{\rm or} ~~~~ \beta_1=8\pi(p + \rho
v^2)/B^2=1 ~, \label{eq:stress balance}
\end{equation}
where $\rho$, $p$, $v$ and  $B$ are the local density, gas
pressure, velocity and magnetic field, respectively. Here,
$\beta_1$ is the generalized plasma parameter, which includes both
\textit{thermal and ram pressure} \citep{RomanovaEtAl2002}. It is
analogous to the standard plasma parameter $\beta=8\pi p/B^2$, but
takes into account the ram pressure of the matter flow in the
disk. Axisymmetric and global 3D MHD simulations show that the
condition $\beta_1=1$  is valuable for finding the magnetospheric
radius $r_m$ (e.g., \citealt{KulkarniRomanova2013}). This radius
$r_m$ corresponds to the innermost edge of the disk, where the
density drops sharply towards the low-density magnetosphere. This
formula can be used to find $r_m$ in many situations, including
those that are non-stationary or where the magnetic field is more
complex than the dipole one. Sometimes, the condition $\beta=8\pi
p/B^2=1$ is used to find the magnetospheric radius (e.g.,
\citealt{BessolazEtAl2008}). This condition, however, yields a
somewhat larger radius at which matter flows from the disk to the
funnel stream.

In theoretical studies  (e.g.,
\citealt{PringleRees1972,LambEtAl1973}) the magnetospheric radius
was estimated from the balance between the largest components of
the stresses, assuming a dipole field of the star and Keplerian
orbital speed in the disk; this gives
\begin{equation}
r_m = k \big[\mu_\star^4/(\dot{M}^2 GM_\star)\big]^{1/7},
~~~~~k\sim 1~, \label{eq:alfven}
\end{equation}
\noindent where $\mu_\star=B_\star R_\star^3$ is the magnetic
moment of the star with a surface field $B_\star$, $\dot{M}$ is
the disk accretion rate, and $M_\star$  and $R_\star$ are the mass
and radius of the star, respectively.

Axisymmetric simulations by, e.g., \citet{LongEtAl2005}  found
that eqs. \ref{eq:stress balance} and \ref{eq:alfven} give similar
values of $r_m$ if $k\approx 0.5$. \citet{KulkarniRomanova2013}
performed a series of 3D MHD simulations of accretion in the
stable regime and compared theoretically-derived values of $r_m$
(using Eq. \ref{eq:alfven}) with those obtained from simulations
(using condition $\beta_1=1$). They found that, in the range of
magnetospheric radii $r_m\sim (2-5) R_\star$, the radius $r_m$
obtained from the simulations can be described by Eq.
\ref{eq:alfven} if $0.55\lesssim k\lesssim 0.72$. Here, we
performed similar comparisons between numerically-derived
magnetospheric radii and Eq. \ref{eq:alfven} using numerical runs
in the stable and unstable regimes (see Appendix
\ref{appendix-rm}).

Another important radius is the corotation radius, $r_{\rm cor}$,
where, by definition, the angular velocity of the star matches the
Keplerian angular velocity of the disc:
$\Omega_\star=\sqrt{GM_\star/r_{\rm cor}^3}$, from which we obtain
$r_{\rm cor}=[GM_\star/{\Omega_\star}^2]^{1/3}$.
 Both the magnetospheric
radius $r_m$ and corotation radius $r_{\rm cor}$ are useful in the
analysis of magnetized stars.

Another convenient parameter that is often used while
investigating the physics at the disk-magnetosphere boundary is
the fastness parameter, which is the ratio between the angular
velocity of the star and the Keplerian angular velocity at the
disk-magnetosphere boundary $r=r_m$ (e.g.,
\citealt{GhoshLamb1978,Ghosh2007}):
\begin{equation}
\omega_s=\frac{\Omega_\star}{\Omega_K(r_m)} ~ ,~~~~~~ {\rm
or,}~~~~~~~ \omega_s=\bigg(\frac{r_m}{r_{\rm cor}}\bigg)^{3/2} .
\label{eq:fastness}
\end{equation}
One can see that this  parameter is also a simple function of the
ratio between  two main radii, $r_m$ and $r_{\rm cor}$. The
fastness parameter $\omega_s$ is expected to be significant in
determining the processes at the disk-magnetosphere boundary
(e.g., \citealt{Ghosh2007}). In our study we use either $\omega_s$
or the ratio $r_m/r_{\rm cor}$.


\begin{figure*}
\centering
\includegraphics[width=10.cm,clip]{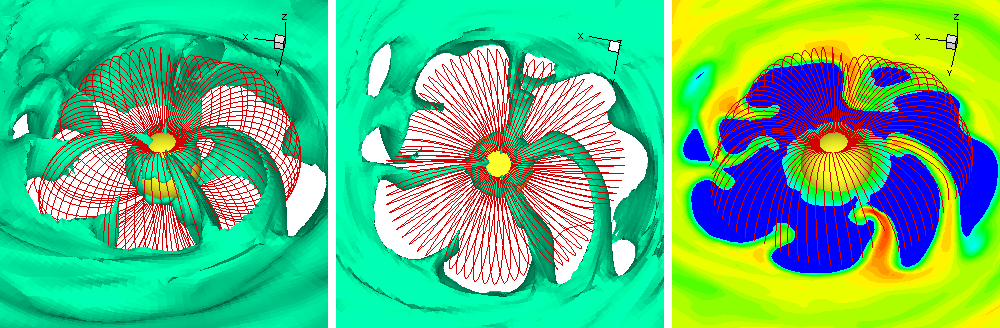}
\caption{\textit{Left panel}: 3D view of matter flow in a case
where chaotic accretion in multiple tongues dominates, model
$\mu1c2.5\Theta5a0.02$, at time $t=19$. One of the density levels
is shown in color, selected magnetic field lines are shown in red.
\textit{Middle panel}: Same but in the face-on projection.
\textit{Right panel:} An equatorial slice of density distribution
is shown in color.}
\label{d1c25t5-3}   
\end{figure*}

\begin{figure*}
\centering
\includegraphics[width=13.3cm,clip]{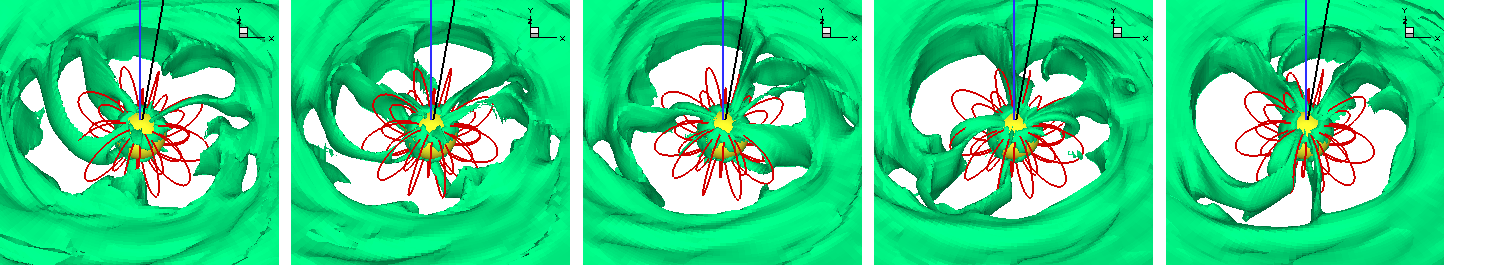}
\includegraphics[width=13.7cm,clip]{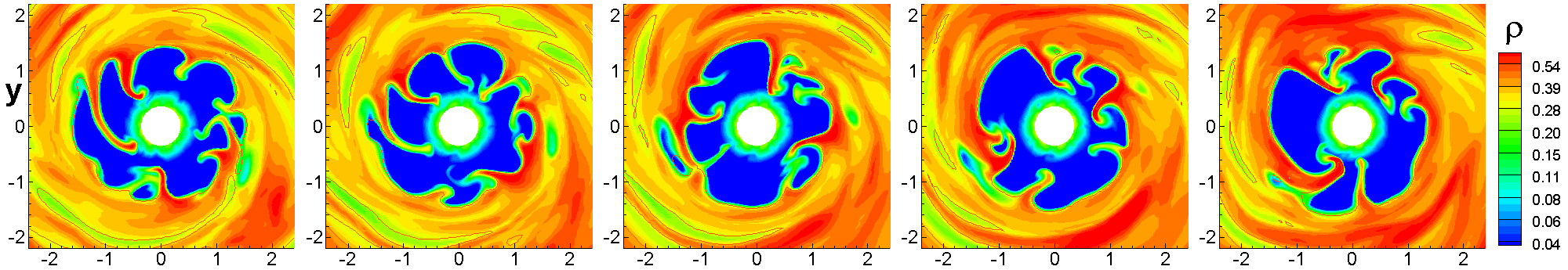}
\includegraphics[width=13.7cm,clip]{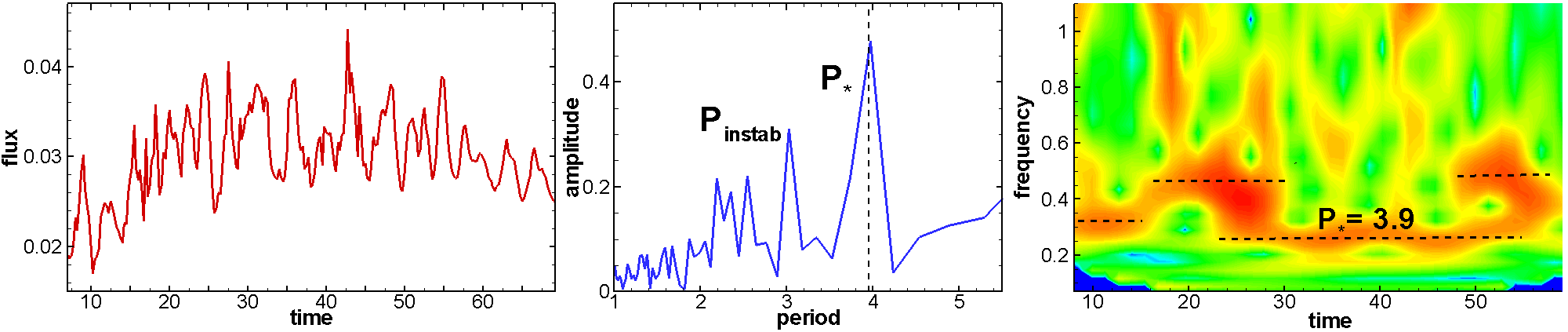}
\caption{An example of accretion in the chaotic unstable regime in
the model $\mu1c2.5{\Theta5}\alpha0.02$ (times $t=19-21$), where
multiple tongues form. Top row shows consecutive 3D views of
matter flow, where the color background represents one of the
density levels and the lines are sample magnetic field lines. The
axes show the directions for the angular momentum and magnetic
momentum of the star. The middle panel shows consecutive xy
slices. The color background shows density distribution and the
lines show where the kinetic plasma parameter $\beta_1=1$. Bottom
row shows the light-curve from rotating spots calculated at an
inclination angle $i=45^\circ$ (left panel), Fourier transform
(middle panel) and wavelet transform (right panel) obtained from
analysis of the light-curve.} \label{d1c25t5-5}
\end{figure*}

\subsection{Rayleigh-Taylor instability at the disk-magnetosphere boundary}

The disk-magnetsophere boundary is prone to Rayleigh-Taylor and
Kelvin-Helmholtz instabilities. Detailed simulations of the
boundary in the shearing box \citep{RastatterSchindler1999} and
also global simulations (e.g., \citealt{KulkarniRomanova2008})
show that the RT instability strongly dominates.
 In a simplified case
 of two adjacent, rotating inviscid fluids with the magnetic field
 perpendicular to the boundary between them, the boundary is unstable if
 the effective gravity
$g_{\rm eff}=g+g_c $ is negative \citep{Chandrasekhar1961}. Here,
$g=-GM_\star/r^2$ and $g_c={\Omega_\star}^2 r$ are gravitational
and centrifugal acceleration, respectively. Taking into account
that at the boundary ($r=r_m$)
 $g=-GM_\star/r^2=-\Omega_K^2 r$, where $\Omega_K=\sqrt{GM_\star/r_m^3}$ is the Keplerian
velocity at the disc-magnetosphere boundary, we obtain
\begin{equation}
g_{\rm eff}=-\Omega_K^2 r + {\Omega_\star}^2 r = - \Omega_K^2 r
(1-\omega_s^2).
\end{equation}
 One can see that in this simplified case the boundary is unstable
 \textit{, if}
 $\omega_s<1$.

\begin{figure*}
\centering
\includegraphics[width=10.0cm,clip]{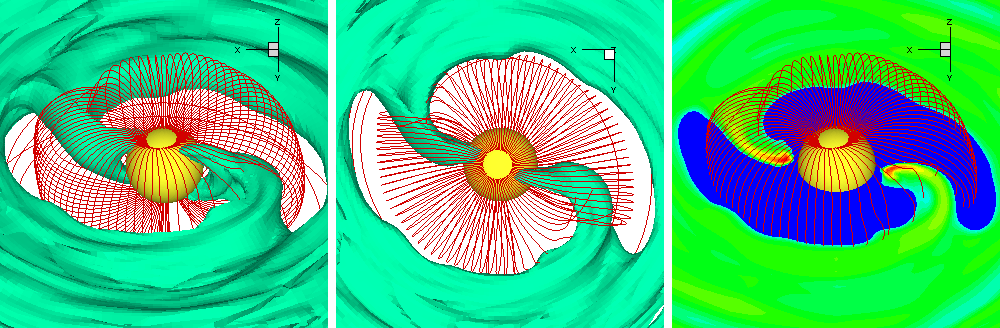}
\caption{\textit{Left panel}: Same as Fig. \ref{d1c25t5-3}, but
for a case where ordered accretion in two tongues dominates, model
$\mu0.5c3\Theta5a0.02$, at time $t=14$.}
\label{d05c3t5-3}     
\end{figure*}

\begin{figure*}
\centering
\includegraphics[width=13.3cm,clip]{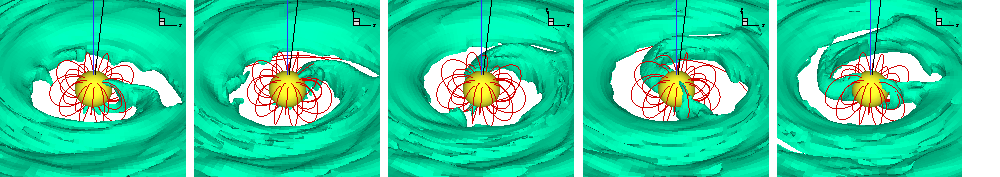}
\includegraphics[width=13.7cm,clip]{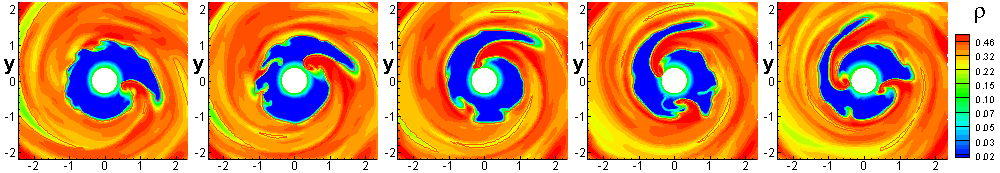}
\includegraphics[width=13.7cm,clip]{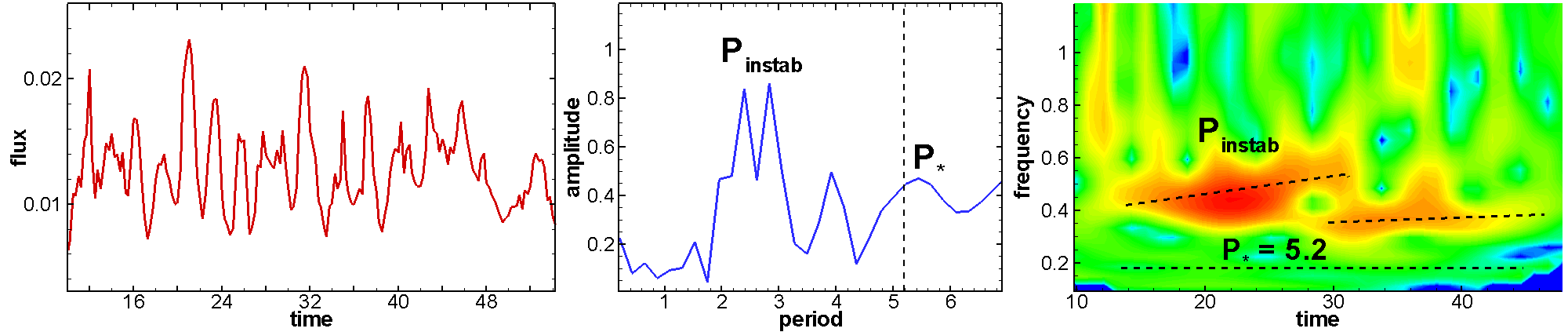}
\caption{Same as Fig. \ref{d1c25t5-5}, but for accretion in the
ordered unstable regime, model $\mu0.5c3\Theta5\alpha0.02$ (times
$t=25.5-27.5$), where one or two ordered unstable tongues form.}
\label{d05c3t5}
\end{figure*}

 A more detailed theoretical analysis shows that in a realistic situation there are a
 number of factors that can suppress instability. One of these factors is viscosity in the disc.
 This viscosity may be associated with the turbulence in the disk.
 Alternatively,
 it can be numerical viscosity associated with the finite size of the grid in numerical models.
 Both types of viscosity can suppress the short wavelength
 perturbations \citep{Chandrasekhar1961}. On the other hand,
 the azimuthal component of the field
 $B_\phi$ can
 also suppress  the short-wavelength perturbation modes
 \citep{Chandrasekhar1961}.
In addition to the above factors, the differential rotation of
matter in the disk, $d\Omega/dr$,
 tends to suppress the RT unstable modes (e.g.,
 \citealt{KaisigEtAl1992}).

 Another important factor that
can increase or decrease the strength
 of instability is the compression factor $K_{\rm
B\Sigma}=\left|\dd{r} \ln \frac{\Sigma}{B_z} \right|$, which is
the gradient of surface density per unit of magnetic flux. The
growth rate of the RT instability is larger if the gradient of
$\Sigma/B_z$  increases fast enough in the direction of gravity
\citep{KaisigEtAl1992, LubowSpruit1995}.

\citet{SpruitEtAl1995} have performed a general analysis of disc
stability in the thin disc approximation, taking the velocity
shear into account The disc has a surface density $\Sigma$ and is
threaded by a magnetic field with the vertical component $B_z$.
Their analytical criterion for the development of instability is:
\begin{equation}
 \gamma_{B\Sigma}^2 \equiv (-g_{\rm eff}) \left| \dd{r} \ln
\frac{\Sigma}{B_z} \right| > 2 \left( r \dd[\Omega]{r} \right)^2
\equiv \gamma_\Omega^2~. \label{eq:spruit}
\end{equation}

One can see that large, negative values of effective gravitational
acceleration
 $g_{\rm eff}$ and large values of the compression factor  $K_{\rm B\Sigma}=\left| \dd{r} \ln \frac{\Sigma}{B_z} \right|$
lead to stronger instability, while the radial shear, $
\gamma_\Omega^2=2 \left( r \dd[\Omega]{r} \right)^2$, is a factor
that suppresses instability by smearing out the perturbations.
This criterion is valuable for the analysis of the
disk-magnetosphere boundary. However, it does not take into
account the possible suppression of perturbation modes by the
azimuthal field, $B_\phi$, or by viscosity. Global 3D MHD
simulations are required to take into account all of these
factors, provided that the grid resolution is fine enough to
resolve (and not suppress) the small-scale perturbation modes.

\begin{figure*}
\centering
\includegraphics[width=16cm,clip]{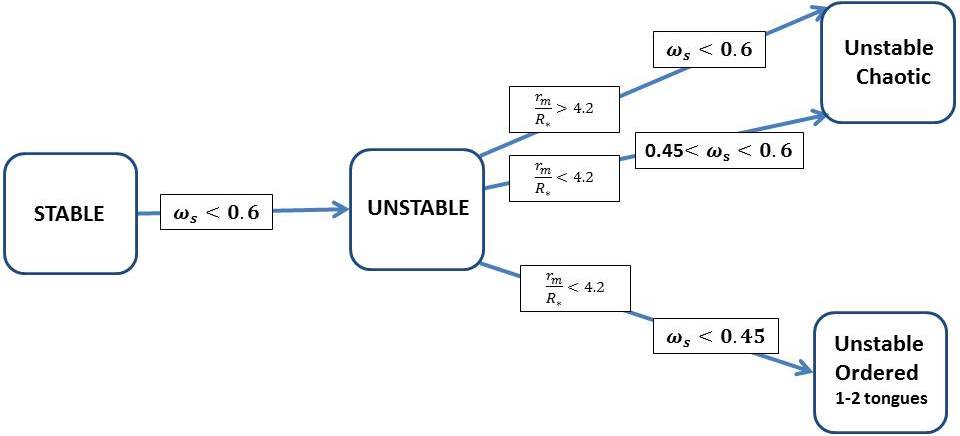}
\caption{The diagram shows transitions between stable and two
types of unstable regimes discussed in the paper and calculated
for low viscosity, $\alpha=0.02$, and a small tilt of the dipole
magnetosphere, $\Theta=5^\circ$. The diagram shows the dependence
of boundaries on the fastness parameter $\omega_s$.   The ratio
$r_m/R_\star$ shows the dimensionless size of the magnetosphere at
which chaotic or ordered type of instability have been observed.}
\label{diagram}
\end{figure*}

\section{Numerical Model}
\label{sec:model}

We perform global 3D MHD simulations of disc accretion onto a
rotating magnetized star. The model we use is the same as in our
earlier 3D MHD simulations of stable and unstable regimes (e.g.,
\citealt{RomanovaEtAl2008, KulkarniRomanova2008}), which has been
described in our earlier papers (e.g.,
\citealt{KoldobaEtAl2002,RomanovaEtAl2003}).  Hence, we will only
describe it briefly here.

\subsection{Initial and boundary conditions}

\textit{Initial conditions.} A star has a dipole magnetic moment
$\mu$, the axis of which makes an angle $\Theta$ with the star's
rotational axis $\Omega_\star$. The rotational axes of the star
and the accretion disc are aligned. A  star is surrounded by an
accretion disc and a corona. The disc is cold and dense, while the
corona is hot and rarefied, and at the reference point (the inner
edge of the disc in the disc plane) the temperature and density
are $T_c=100T_d$, and $\rho_c=0.01\rho_d$, where subscripts `d'
and `c' denote the disc and the corona. Initially, the disc and
corona are in rotational hydrodynamic equilibrium (see e.g.,
\citealt{RomanovaEtAl2002} for details). The disc is relatively
thin, with the half-thickness to radius ratio $h/r\approx 0.1$.

\smallskip

\noindent\textit{Boundary conditions.} At both the inner and outer
boundaries, most of the variables $A_j$ are taken to have free
boundary conditions, ${\partial A_j}/{\partial r}=0$. The free
boundary conditions on the hydrodynamic variables at the stellar
surface mean that accreting gas can cross the surface of the star
without creating a disturbance in the flow. These conditions
neglect the complex physics of interaction between the accreting
gas and the star \footnote{The impact of accreting matter with the
stellar surface and  formation of the radiative shock wave near
the surface has been  investigated, e.g.,  by
\citet{KoldobaEtAl2008,OrlandoEtAl2010}.}. The magnetic field is
frozen onto the surface of the star. That is, the normal component
of the field, $B_n$, is fixed.
    The other components of the
magnetic field vary. At the outer boundary, matter is not
permitted to flow into the region. The simulation region is
usually large enough for the disc to have enough mass to sustain
accretion flow for the duration of the simulation run.

\subsection{MHD equations, numerical method and the grid}

 The full set of 3D MHD equations taken in
dimensionless form is solved numerically using a Godunov-type
numerical scheme, written in a ``cubed-sphere'' coordinate system
which rotates with the star \citep{KoldobaEtAl2002}. The numerical
approach is similar to that described in \citet{PowellEtAl1999},
where the seven wave Roe-type approximate Riemann solver is used.
The energy equation is written in the form of entropy balance, and
the equation of state is that of an ideal gas. Compared with
\citet{KoldobaEtAl2002}, viscosity terms are incorporated into the
equations. Viscosity is modelled using the $\alpha$-model
\citep{ShakuraSunyaev1973}, and is incorporated only into the
disc, so that it controls the accretion rate through the disc. In
contrast with our earlier studies, we use a small
$\alpha$-parameter $\alpha=0.02$ in most of the simulation runs,
and use a larger value $\alpha=0.1$ in the test runs.

A ``cubed sphere" grid consists of $N_r$ concentric spheres, where
each sphere represents an inflated cube. Each of the six sides of
the inflated cube has an $N\times N$ curvilinear grids which
represent a projection of the Cartesian grid onto the sphere. The
entire grid consists of $6 N^2\times N_r$ cells. The typical grid
used in our simulations has $N_r=140$ cells in the radial
direction, and $N^2 = 61^2$ \textit{`angular'} cells in each
block. This grid is twice as fine as the $N^2 = 31^2$ grid used in
our earlier studies of instabilities (e.g.,
\citealt{KulkarniRomanova2008}). To check the convergence of
results at higher grid resolutions, the following grids were also
calculated:
 $51^2\times110$,
$71^2\times160$, $81^2\times185$, and $101^2\times220$.
Simulations show that at a small parameter of viscosity
($\alpha=0.02$) the coarsest grid, $31^2$, suppresses instability,
while at all finer grids the results do not depend on the grid
resolution (see details in Sec. \ref{sec:grids}).

\subsection{Dimensionalization}

 Equations are written using
\textit{dimensionless variables}. The dimensionless value of every
physical quantity $A_j$ is defined as $\tilde{A_j} = A_j/A_{j0}$,
where $A_{j0}$ is the reference value for $A_j$. The simulations
were performed in dimensionless variables $\widetilde{A}$.

First, we define the main reference values: the reference scale,
$R_0=R_\star/0.35$, where $R_\star$ is the radius of the star
\footnote{This reference scale has been used in our prior models
(e.g., Koldoba et al. 2002; Romanova et al. 2002), and we now use
it for consistency with our earlier works.}; the reference mass,
$M_0=M_\star$, where $M_\star$ is the mass of the star; and the
magnetic moment of the star, $\mu_\star=B_\star R_\star^3$, where
$B_\star$ is the magnetic field at the magnetic equator. Then, we
determine the other reference values, which include: velocity
$v_0=(GM_0/R_0)^{1/2}$;
 time-scale  $t_0=R_0/v_0$; period of rotation at $r=R_0$: $P_0=2\pi R_0/v_0$ (
 used in our plots); reference angular velocity
$\Omega_0=v_0/R_0$; and reference frequency $f_0=\Omega_0/2\pi$.
 We determine the reference magnetic
field $B_0$ and magnetic moment $\mu_0=B_0 R_0^3$ such that
$\mu_0=\mu_\star/\widetilde\mu$, where $\widetilde\mu$ is the
dimensionless parameter used to vary the size of the dimensionless
magnetosphere. The reference field is
$B_0=\mu_\star/(R_0^3\widetilde\mu)$, the reference density is
$\rho_0=B_0^2/v_0^2$, the reference mass accretion rate is
$\dot{M}_0=\rho_0 v_0 R_0^2 = \mu_\star^2/(\widetilde{\mu}^2 R_0^4
v_0)$, and the reference surface density is $\Sigma_0=\rho_0 R_0$.

The results obtained in dimensionless variables $\widetilde{A}$
can be applied to different types of stars. We determine the
dimensional mass, radius and magnetic field of a star and derive
other reference values, as discussed above. Table \ref{app:refval}
lists the reference values for three classes of stars: classical T
Tauri stars, white dwarfs and neutron stars.
 To obtain the physical dimensional values $A$, the dimensionless
values $\widetilde{A}$ should be multiplied by the corresponding
reference values $A_0$ as $A=\widetilde{A}A_0$.
 Subsequently, in the text we drop the tildes above the dimensionless variables
and show dimensionless values everywhere unless otherwise
specified.

\subsection{Main dimensionless parameters of the model.}
\label{sec:main-dim-parameters}

Taking into account the above dimensionalization,
 the number of parameters in the problem is relatively small.
The dimensionless mass and radius of the star, $\tilde M_\star=1$
and $\tilde R_\star=0.35$, respectively, are fixed throughout all
simulation runs. We fix the fiducial density values in the disk
and corona at
 $\tilde\rho_d=1$ and $\tilde\rho_c=0.01$, respectively,
 and use the same disk structure  across all simulation runs.
The only parameter that is varied is the $\alpha-$parameter of
viscosity, which regulates the radial velocity in the disk,
$v_r\sim\alpha$ \citep{ShakuraSunyaev1973}. We use a small
$\alpha-$parameter,
 $\alpha=0.02$, in most simulation runs.

There are two main dimensionless parameters that we vary. One of
them is the dimensionless corotation radius $\tilde r_{\rm cor}$,
which determines the dimensionless angular velocity of the star,
$\tilde \Omega_\star=\tilde r_{\rm cor}^{-3/2}$. This is one of
the key parameters in determining the boundary between stable and
unstable regimes of accretion.

The second important parameter is the dimensionless magnetic
moment $\widetilde\mu$, which determines the size of the
dimensionless magnetosphere, $\widetilde r_m =r_m/R_0$. We do not
know the size of the magnetosphere in advance. Instead, we obtain
it from the simulations. We can estimate the expected size of the
magnetosphere using the theoretical formula  Eq. \ref{eq:alfven}
and substituting the magnetic moment of the star,
 $\mu_\star$, with that obtained from our dimensionalization
 formalism:   $\dot{M}_0=
\mu_\star^2/(\widetilde{\mu}^2 R_0^4 v_0)$, or
$\mu_\star^2=\dot{M}_0 \widetilde{\mu}^2 R_0^4 v_0$ and
$v_0=(GM/R_0)^{1/2}$. We obtain:

\begin{equation}
\widetilde r_m=r_m/R_0=k
({\widetilde\mu}^2/\widetilde{\dot{M}})^{2/7}~.
\label{eq:rA-dimensionless}
\end{equation}
Here,  $\widetilde{\dot{M}}$ is the dimensionless accretion rate,
which we find from simulations and typically does not vary much
across simulations with similar $\alpha-$parameters of viscosity.
The main parameter that determines the dimensionless size of the
magnetosphere, $r_m/R_\star$, is parameter $\widetilde\mu$. These
two parameters ($\widetilde r_{\rm cor}$ and $\widetilde\mu$)
determine the main physics at the disk-magnetosphere boundary
\footnote{Note that the coefficient $k$ equals to that in Eq.
\ref{eq:alfven}.}. In the following sections, we drop the tilde's
above $\widetilde\mu$ and $\widetilde r_{\rm cor}$ for
convenience.

The results also depend on the tilt of the dipole magnetosphere,
$\Theta$. We investigated the cases of small ($\Theta=5^\circ$)
and larger ($\Theta\geq 20^\circ$) tilts in separate sets of runs.

\begin{figure}
\centering
\includegraphics[width=8.5cm,clip]{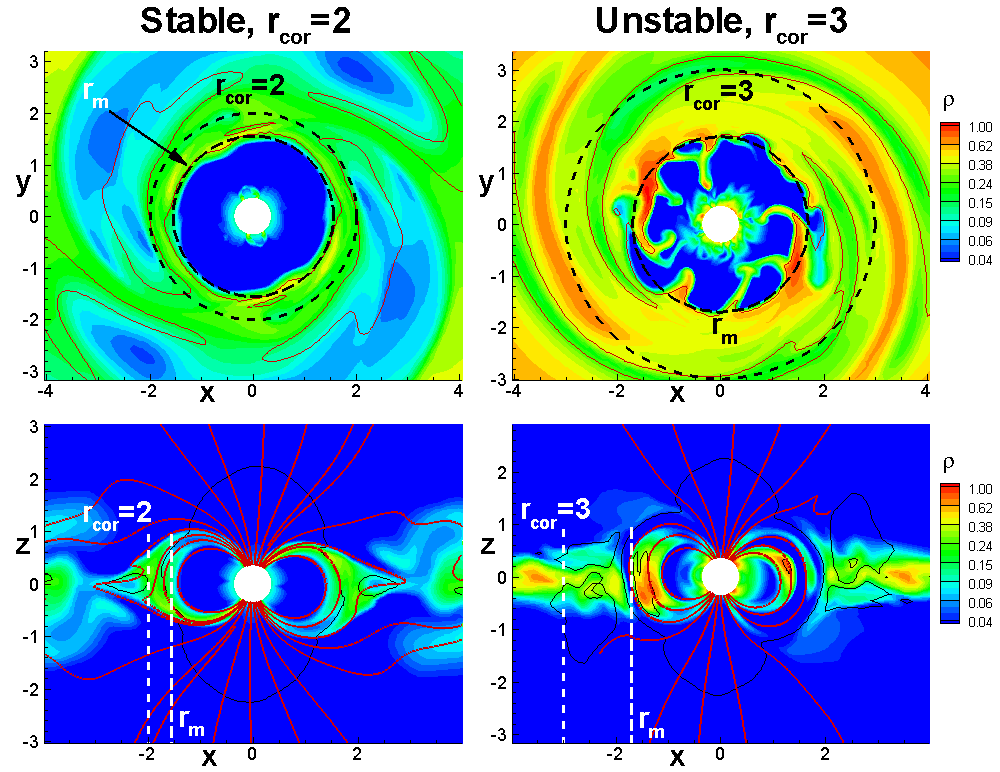}
\includegraphics[width=8.5cm,clip]{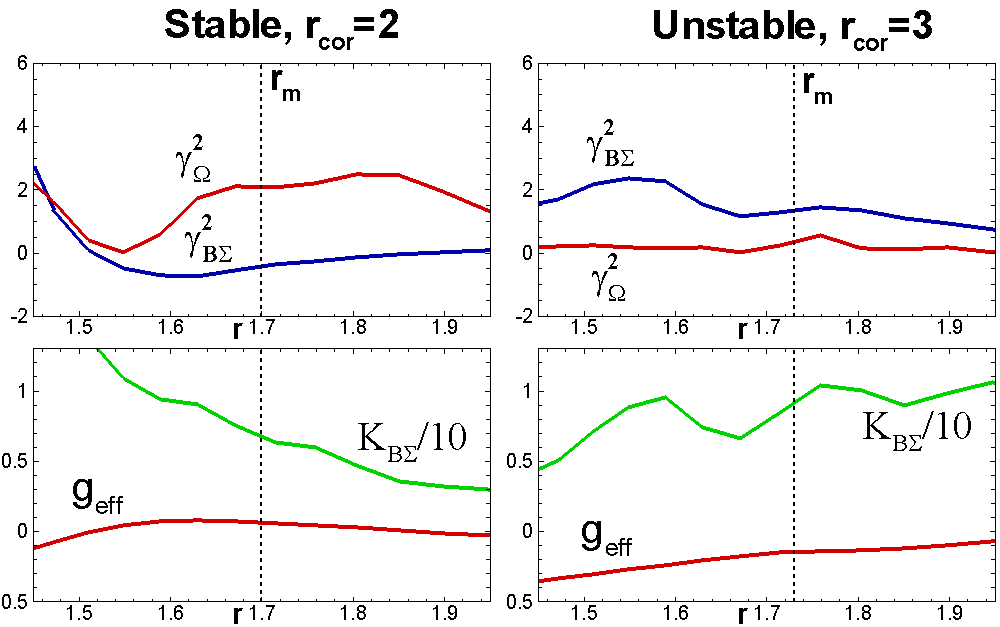}
\caption{\textit{Top four panels:} Example of two simulation runs
with same parameters ($\mu=2$, $\Theta=5^\circ$, $\alpha=0.02$ at
time $t=20$) but different  corotation radii (models
$\mu2c2{\Theta5}\alpha0.02$ and $\mu2c3{\Theta5}\alpha0.02$).
\textit{Left panels}: xy-slice (top) and xz-slice (bottom) show
the density distribution in case of $r_{\rm cor}=2$, where
accretion is stable. \textit{Right panels}: accretion becomes
unstable when $r_{\rm cor}=3$. Red lines show sample magnetic
field lines. Positions of the magnetospheric $r_m$ and corotation
radii are shown with the dashed lines. \textit{Bottom four
panels:} Radial distribution of of different terms of the Spruit
criterion (Eq. \ref{eq:spruit}) in the vicinity of $r_m$ at time
$t=6$ (in the beginning of the unstable regime) for models shown
above. } \label{d2-stable-unstable-2}
\end{figure}

\begin{figure*}
\centering
\includegraphics[width=6.cm]{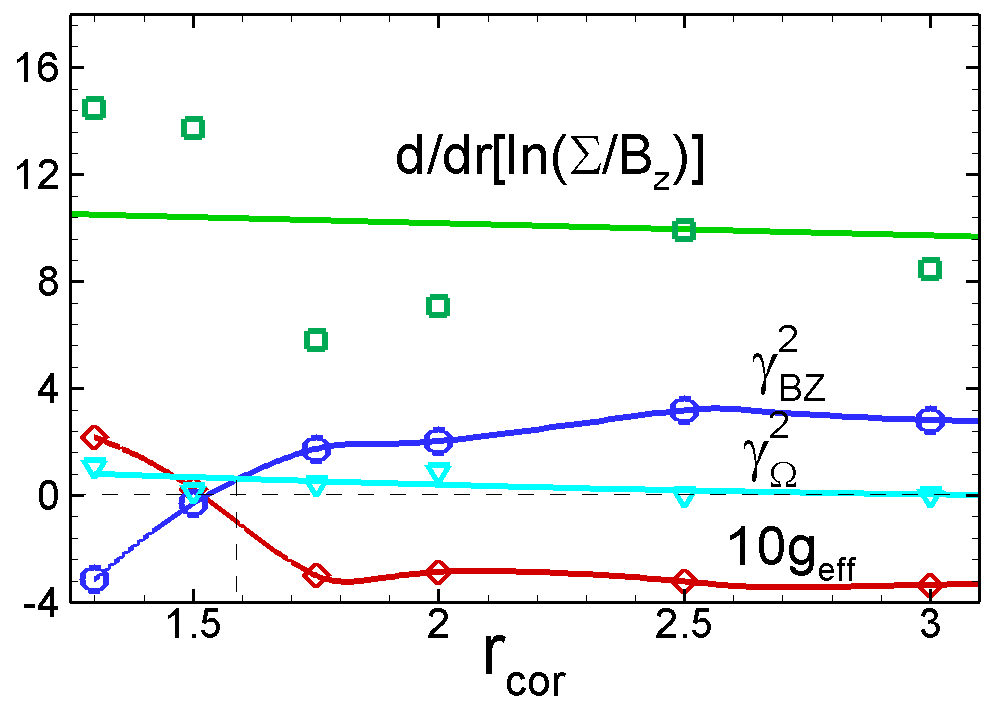}
\includegraphics[width=6.cm]{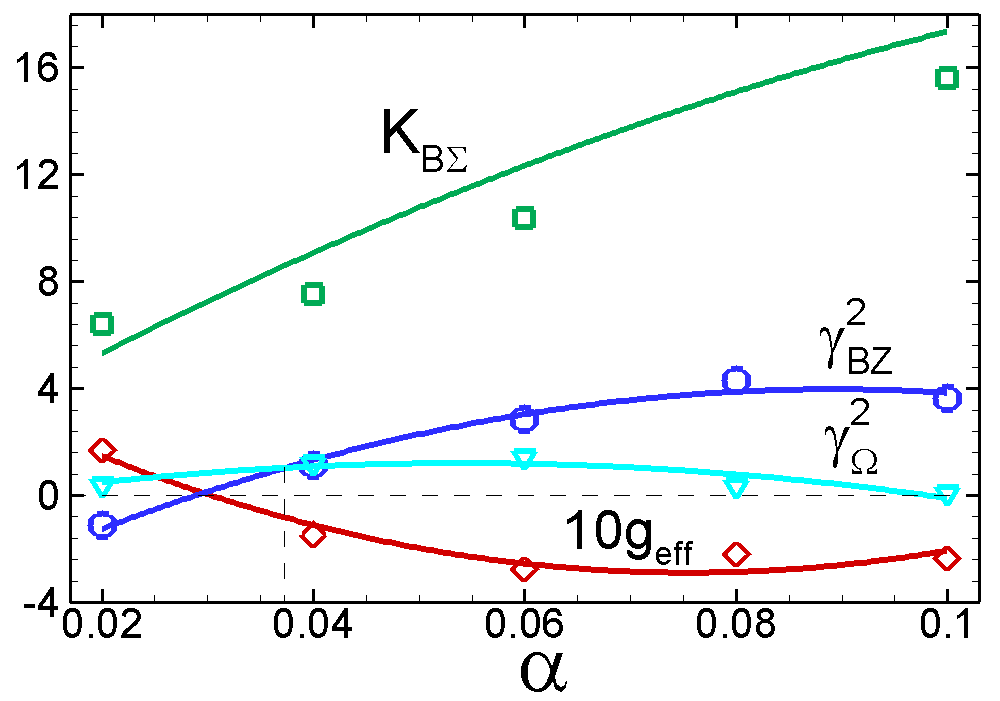}
\caption{\textit{Left Panel:} Dependence of different terms from
Spruit's criterion on corotation radius $r_{\rm cor}$. Simulations
were performed for stars with $\mu=1$, $\alpha=0.02$ and different
$r_{\rm cor}$. $g_{\rm eff}$ is multiplied by 10 for clarity.
\textit{Right Panel:} Same but for dependence on $\alpha-$
parameter of viscosity. Simulations were done for $\mu=1$, $r_{\rm
cor}=1.5$ and different $\alpha-$parameters of viscosity.}
\label{sigav-alpha-r_cor}
\end{figure*}

\section{Boundary Between Stable and Unstable
Regimes of Accretion}

One of the main goals of present research was to find a set of
parameters which would determine the boundary between stable and
unstable regimes of accretion.

\subsection{Set of performed simulations}

To find the boundary between stable and unstable regimes of
accretion, we performed a series of simulation runs, where we kept
the same initial conditions for the density and pressure
distribution in the disc and corona, but varied the the main
dimensionless parameters of the model: magnetic moment $\mu$ of
the star (which determines the dimensionless size of the
magnetosphere) and the corotation radius $r_{\rm cor}$ (which
determines the period of the star). We used a small parameter of
viscosity, $\alpha=0.02$, in the majority of simulation runs to be
sure that viscosity is not an essential factor in generating or
suppressing instability.

We ran two main sets of simulations for two different misalignment
angles of the dipole moment: one at a relatively small angle,
$\Theta=5^\circ$, where instability is expected to be stronger,
and the other at a larger angle, $\Theta=20^\circ$, where funnel
stream accretion is expected to be more favorable (e.g.,
\citealt{KulkarniRomanova2008}).

For each angle $\Theta$, we performed a series of simulations for
stars with different corotation radii in the range of $r_{\rm
cor}=1.2-5$ and parameters $\mu$ ranging from $\mu=0.1$
(small-sized magnetospheres) to $\mu=3$ (large-sized
magnetospheres).
 Table \ref{tab:models} shows
sample models used for finding the boundary. Many of these models
were also used for detailed analysis. The names of the models in
the Table incorporate the parameters used in those models.

To check the main hypothesis (which is the dependence of the
boundary on the fastness parameter $\omega_s$ and hence the ratio
$r_m/r_{\rm cor}$), we measured $r_m$ in each simulation run and
marked whether accretion was stable or unstable at a given value
of $r_{\rm cor}$ (see Fig. \ref{boundary-t5-t20}).

The accretion rate and $r_m$ generally vary in time. In some
cases, accretion is either stable or unstable throughout the
entire simulation run, while in other cases it is only marginally
stable, and can transition from stable to unstable and back to
stable again. In each case, we measured $r_m$ from the simulations
using the position of the $\beta_1=1$ line for a few moments in
time, and marked whether the case was stable or unstable. The
$\beta_1=1$ line is not a smooth line: it reflects the inner parts
of the disc as well as the matter-dominated unstable tongues. We
determined $r_m$ using only the inner disc radius and ignoring the
tongues. All of these points were used for finding the boundary
between stable and unstable regimes of accretion.

\subsection{Results}
\label{sec:results}

Fig.~\ref{boundary-t5-t20} (left panel) shows the results of
simulations in the $r_m - r_{\rm cor}$ parameter space for a small
misalignment angle of the dipole, $\Theta=5^\circ$. We can draw an
approximate boundary line between stable (red squares) and
unstable (blue triangles and green x's) regimes of accretion. This
line corresponds to the ratio $r_m/r_{\rm cor}\approx 0.71$, or
the fastness parameter $\omega_s\approx 0.6$ (see bold line in the
figure). The plot shows that accretion is stable above this line,
and is unstable otherwise. In this figure, the values of $r_m$ and
$r_{\rm cor}$ are given in radii of the star for user convenience.
As expected, instability occurs more easily when  $r_m$ is smaller
and $r_{\rm cor}$ is larger, that is, when the magnetosphere
rotates more slowly than the inner disc and the fastness parameter
$\omega_s$ is smaller.

At a sufficiently small ratio of $r_m/r_{\rm cor}$, another
transition occurs and matter starts accreting in the
\textit{ordered unstable regime}. The boundary between chaotic and
ordered unstable regimes approximately corresponds to the
condition of $r_m/r_{\rm cor}\approx 0.59 $, or fastness parameter
$\omega_s\approx 0.45$. This new regime has only been observed in
the cases of relatively small magnetospheres, $r_m/R_\star\lesssim
4.2$. At larger magnetospheres, $4.2 \lesssim r_m/R_\star\lesssim
7$, accretion is chaotic (see left panel of
Fig.~\ref{boundary-t5-t20}).

We should note that at even larger magnetospheres, $r_m\gtrsim 7
R_\star$, the RT instability only develops in the external regions
of the magnetosphere (e.g.,
\citealt{RomanovaEtAl2014,RomanovaOwocki2015}) and accretion in
two funnel streams dominates. The parameter space with large $r_m$
should be investigated separately

We should also note that when $r_m/r_{\rm cor}>1$ (or, fastness
$\omega_s>1$), the magnetosphere rotates more rapidly than the
inner disc, and the propeller regime is expected. We observed from
the simulations that when $r_m\approx r_{\rm cor}$ (fastness
parameter $\omega_s=1$), the magnetosphere pushes the disc outward
through the propeller mechanism. The parameter values under which
the propeller regime is expected to dominate should be studied
separately. In the present study, we draw the $\omega_s=1$ line as
a reminder that the propeller regime is expected near or above
this line.

\begin{figure*}
\centering
\includegraphics[width=13.cm,clip]{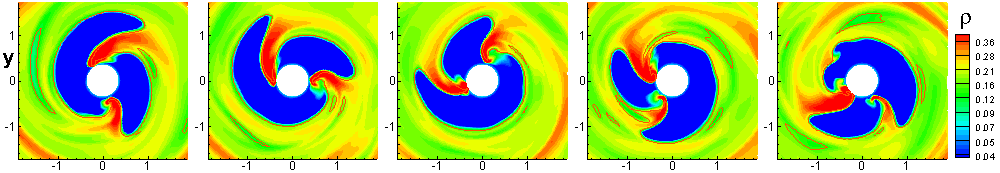}
\includegraphics[width=13cm,clip]{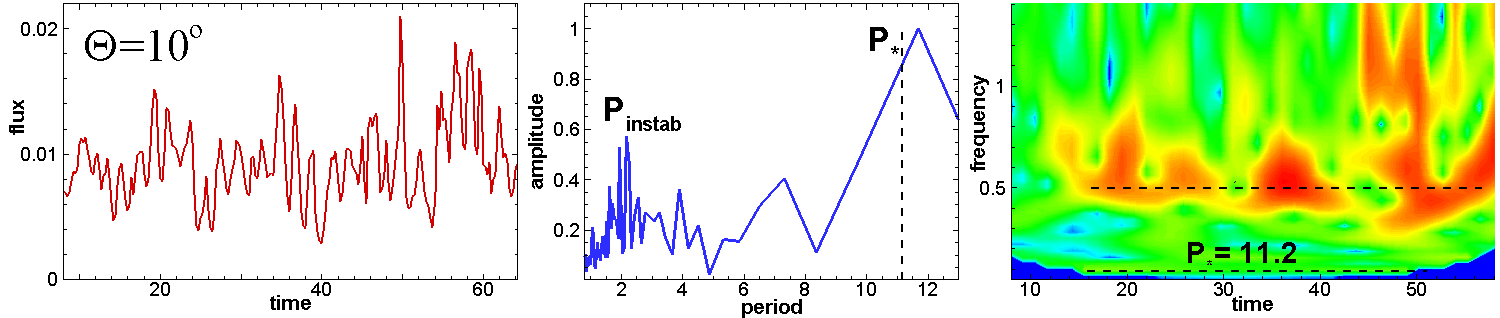}
\includegraphics[width=13.cm,clip]{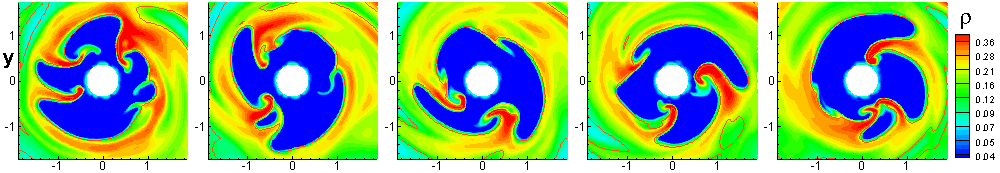}
\includegraphics[width=13cm,clip]{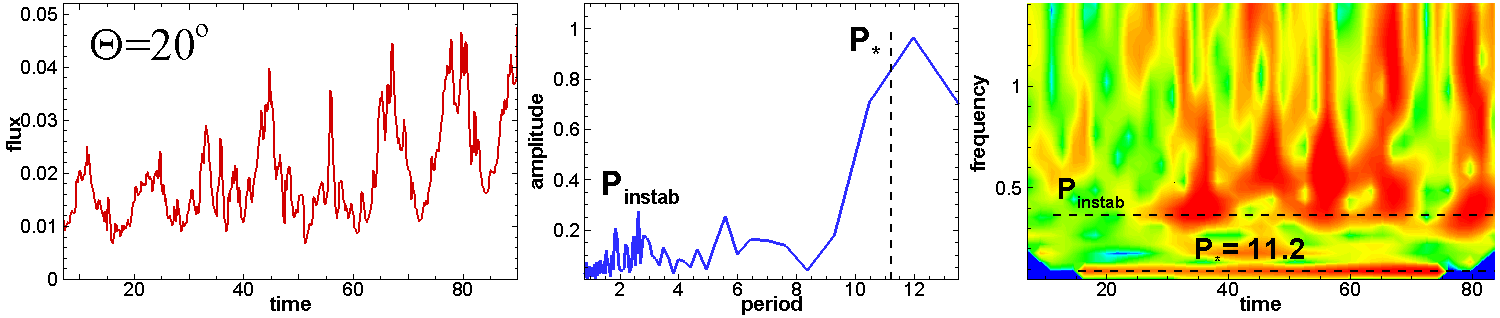}
\caption{\textit{Top two rows:} consecutive xy-slices (times
$t=37-39$) and analysis of variability for a larger misalignment
angle, $\Theta=10^\circ$, model $\mu0.3c5\Theta10\alpha0.02$.
\textit{3rd and 4th rows:} same as top two rows, but for
$\Theta=20^\circ$, model $\mu0.5c5\Theta20\alpha0.02$ (times
$t=30.75-32.75$).}
\label{theta-diff-6}
\end{figure*}

\section{Two types of unstable regime: Chaotic and Ordered}
\label{sec:ordered-chaotic}

Here, we describe the two types of unstable regime in greater
detail, as observed in the simulations: (1) the \textit{chaotic
unstable regime}, where  matter accretes in several chaotic,
transient tongues, and (2) the \textit{ordered unstable regime},
where one or two tongues form and rotate persistently with the
frequency of the inner disc. Below, we discuss these two types of
unstable regime for cases of a small tilt of the dipole,
$\Theta=5^\circ$.

\subsection{Chaotic unstable regime}
\label{sec:unstable-chaotic}

To demonstrate the chaotic unstable regime, we use the model
$\mu1c2.5\Theta5\alpha0.02$, where the magnetosphere is relatively
large ($\mu=1$) and the corotation radius is $r_{\rm cor}=2.5$,
which corresponds to the period $P_\star\approx 3.9$ in
dimensionless units. We observed that matter accretes in several
tongues, which form frequently and rapidly disappear. Fig.
\ref{d1c25t5-3} (left two panels) shows a typical picture of
matter flow through the chaotic unstable tongues, which are tall
and narrow, and which penetrate the field lines by pushing them
aside (right panel).

Fig. \ref{d1c25t5-5} shows the temporal evolution of tongues in
several consecutive slices of density distribution, separated by
the time interval of $\Delta t=0.5$ \footnote{Note that in our
units $\Delta t=1$ corresponds to the period of Keplerian rotation
at $r=1$. However, the inner disc usually rotates with
sub-Keplerian velocity, because the disc interacts with the slowly
rotating magnetosphere}. Top two rows show that the tongues are
constantly modified, breaking or coalescing on a time-scale of
$\Delta t \approx 0.5-1$. The $3D-$panels show accretion through
chaotic instabilities, with the occasional formation of funnel
streams.

The matter from unstable tongues accelerates due to gravity, falls
onto the star and releases kinetic energy at the surface of the
star, forming hot spots. We use a simple model for the radiation
of the spots, suggesting that all kinetic energy of the falling
matter is converted into radiation, which is distributed
isotropically (see details in \citealt{RomanovaEtAl2004}). We
calculate the light-curve as seen by a remote observer, located at
an angle of $i=45^\circ$ with respect to the rotational axis of
the star (see bottom left panel of Fig. \ref{d1c25t5-5}). The
light-curve shows irregular variations on a time-scale $\approx
1.5$ times shorter than the period of the star, $P_\star=3.9$. The
bottom middle and right panels show the Fourier and wavelet
spectra of this light-curve. The wavelet shows different angular
frequencies. One of them corresponds to the frequency of the star,
$f_\star=0.26$ (corresponding to the period of $P_\star=3.9$).
Other frequencies are higher, with $f_{\rm inst}\approx 0.3, 0.5$.
The Fourier spectrum shows that there are several peaks in the
interval of $P_{\rm inst}=2-3.2$, which correspond to the wavelet
frequencies. These periods also correspond to the time-scale of
variability observed in the light-curve. We suggest that these
periods reflect the frequency of formation of the strongest
tongues that reach the surface of the star.

Both the wavelet and the Fourier analysis show the presence of the
stellar period, which can be associated with the fact that the
entire set of unstable temporary spots is oriented around the
magnetic pole. The magnetic pole is slightly tilted about the
rotational axis (at $\Theta=5^\circ$ in this case) and the
rotation of the star modulates the light from the set of chaotic
spots. This is a probable reason for why the stellar period is
also observed. Another possible reason is that some matter
accretes to the star in funnel streams above the magnetosphere and
forms hot spots which tend to rotate with the star. In the present
study, we chose an inclination angle of $i=45^\circ$ because this
is where modulation by stellar rotation is expected to be the
strongest. If this angle were smaller, then the stellar period
would have had a smaller amplitude.

\subsection{Ordered unstable regime: one- and two-tongue accretion}
\label{sec:unstable-ordered}

When the ratio $r_m/r_{\rm cor}$ becomes sufficiently small,
unstable accretion becomes ordered: multiple tongues merge,
forming one or two ordered tongues. In some cases, two tongues
carry a comparable amount of matter flux, while in other cases one
of the two tongues may carry more mass. In all cases, the bases of
the tongues rotate with the frequency of the inner disc.

Fig. \ref{d05c3t5-3} shows a snapshot of unstable accretion in two
tongues, model $\mu0.5c3\Theta5\alpha0.02$. One can see that two
ordered matter-dominated tongues push the magnetic field lines
apart and penetrate into the deep layers of the magnetosphere,
where they transition into regular funnel streams, but very close
to the star. The top two panels of Fig. \ref{d05c3t5} show
snapshots of accretion at five consecutive moments in time. The
figure shows that initially there was only one tongue, and the
second one formed later. One and two-tongue accretion modes are
often observed in the same simulation run.

The tongues deposit matter onto the stellar surface and form one
or two hot spots that rotate more rapidly than the star. The
light-curve shows regular peaks that reflect the frequency of
rotation of these spots (not the frequency of stellar rotation).
The Fourier analysis shows a large peak associated with the
rotating spots, with a period of $P_{\rm inst}\approx 2-3$. The
period of the star, $P_\star=5.2$, is also visible, but with a
smaller amplitude. The wavelet analysis shows frequencies in the
range of $f_{\rm inst}\approx 0.3-0.6$, which correspond to the
peaks observed in the Fourier spectrum.

The main period observed in the light-curve is associated with the
rotation of the ordered unstable tongues and approximately
corresponds to the period of inner disc rotation. The ordered
unstable regime may be an important mechanism in the formation of
quasi-periodic oscillations during episodes of enhanced accretion,
when the magnetosphere is compressed and the magnetospheric radius
can be much smaller than the corotation radius.

\subsection{The boundary between
chaotic and ordered unstable regimes} \label{sec:unstable-chaotic}

To find the boundary between the chaotic and ordered unstable
regimes, we marked the cases in which one or two unstable tongues
carry more mass than the other tongues and labelled these models
as green x's in Fig. \ref{boundary-t5-t20}. The approximate
position of this boundary corresponds to the  ratio $r_m/r_{\rm
cor}\approx 0.59$ (the fastness $\omega_s\approx 0.45$ in the
cases of relatively small magnetospheres, $r_m/R_\star\lesssim
4.2$. Ordered unstable accretion in one or two tongues dominates
below this line. In the cases of larger-sized magnetospheres
($r_m/R_\star\gtrsim 4.2$), only the chaotic unstable regime has
been observed.

Simulations show that near the $r_m/r_{\rm cor}\approx 0.59$ line
accretion is mainly chaotic, although at some point in a
simulation run one or two tongues may start to carry more matter
than the other tongues. Accretion may also alternate between
ordered and chaotic within the same simulation run. At smaller
values of $r_m/r_{\rm cor}$ accretion becomes systematically more
ordered, and the frequency associated with the ordered tongues
becomes stronger. This means that at smaller values of the
$r_m/r_{\rm cor}$ ratio the quality factor $Q=f/{\Delta f}$
\footnote{The quality factor is the ratio between the frequency of
oscillations and the width of the peak at the half-maximum.} of
the quasi-periodic oscillations associated with the ordered
unstable tongues is higher. Such an increase in the quality factor
is expected during accretion outbursts, when the disc moves inward
towards the star.

The diagram in Fig. \ref{diagram} summarizes the boundaries
between the stable and unstable regimes, and between the chaotic
and ordered unstable regimes.

\section{Analysis of instability}
\label{sec:grids}

In this section, we analyze the dependence of the transition
between stable and unstable regimes on different parameters. we
are interested in understanding the dependence of accretion mode
on the corotation radius $r_{\rm cor}$ and the $\alpha-$parameter
of viscosity.

\subsection{Comparison of stable and unstable cases}
\label{sec:comp-stab-unstab}

To compare the simulation results with the theoretical criterion
(Eq. \ref{eq:spruit}), we took two models with the same parameters
but different corotation radii: $\mu2c2\Theta5\alpha0.02$ and
$\mu2c3\Theta5\alpha0.02$, and compared their modes of accretion.
In the first model, the corotation radius is $r_{\rm cor}=2$, and
a star is in stable regime of accretion (see left panels of
Fig.~\ref{d2-stable-unstable-2}). In the second model, the
corotation radius is $r_{\rm cor}=3$, the star rotates more
slowly, and the mode of accretion is unstable. (see right panels
of Fig.~\ref{d2-stable-unstable-2}). We then calculated the
different terms of the theoretical criterion  (Eq.
\ref{eq:spruit}) in the vicinity of the magnetospheric radius
$r_m$, where the unstable perturbations can occur. To calculate
these terms, we took the azimuthally-averaged values of relevant
variables and plotted them as a function of radius in Fig.
\ref{d2-stable-unstable-2}. Top panels show that in the stable
regime (left panel, $r_{\rm cor}=2$), $\gamma_\Omega^2
> \gamma_{B\Sigma}^2$ at the magnetospheric radius (vertical
dashed line), while in the unstable regime (right panel, $r_{\rm
cor}=3$), $\gamma_{B\Sigma}^2 > \gamma_\Omega^2$. In both cases,
results of the simulations are consistent with the analytical
prediction (Eq. \ref{eq:spruit}). Bottom panels show that in the
unstable regime the effective gravity is $g_{\rm eff}\approx
-0.15$, while in the stable regime it is slightly positive,
$g_{\rm eff}\approx 0.06$.

The compression factor $K_{\rm B\Sigma}$ is larger in the unstable
regime. Therefore, the term characterizing the instability,
$\gamma_{B\Sigma}^2$, is larger in the unstable regime due to both
a larger effective gravity term $-g_{\rm eff}$ and a larger
compression factor $K_{\rm B\Sigma}$.

\subsection{Dependence of instability on $r_{\rm cor}$}

To better understand the physics of transition between the stable
and unstable regimes and its dependence on the $r_{\rm cor}$
parameter, we take the model $\mu1c1.5\Theta5\alpha0.02$ (which is
at the boundary between stable and unstable regimes) and
recalculate it at different corotation radii $r_{\rm cor}$ while
keeping all the other parameter values fixed ($\mu=1$,
$\Theta=5^\circ$, $\alpha=0.02$). We then calculate the main terms
from Eq. \ref{eq:spruit}, $\gamma_{B\Sigma}^2$ and
$\gamma_\Omega^2$, which determine whether the model is stable or
unstable. We also separately calculate the compression factor,
$K_{\rm B\Sigma}=\left| \dd{r} \ln \frac{\Sigma}{B_z} \right|$,
and the effective gravity term, $g_{\rm eff}$.

Fig. \ref{sigav-alpha-r_cor} (left panel) shows that effective
gravity $g_{\rm eff}$ decreases with $r_{\rm cor}$, changes sign
from positive to negative, and then levels off at some negative
value. It levels off because at large corotation radii the angular
velocity is small and the centrifugal acceleration term in $g_{\rm
eff}$ becomes negligibly small compared with the gravitational
acceleration. One can also see that the compression factor $K_{\rm
B\Sigma}$ is large, but does vary systematically, and therefore
the variation of the instability term $\gamma_{B\Sigma}^2$ is
mainly determined by $g_{\rm eff}$. We should note that, although
both $g_{\rm eff}$ and the shear term $\gamma_\Omega^2$ are
relatively small and do not vary much with $r_{\rm cor}$, $g_{\rm
eff}$ becomes important due to its critical sign change during the
transition. We conclude that the transition from the stable to the
unstable regime is largely due to the variation in effective
gravity. The transition occurs at $r_{\rm cor}\approx 1.6$, when
$\gamma_{B\Sigma}^2 = \gamma_\Omega^2$.

We should note that the point of transition, $r_{\rm cor}\approx
1.6$, is close to the value of $r_{\rm cor}\approx 1.5$, where
$g_{\rm eff}$ changes sign from slightly positive to increasingly
more negative.

\begin{figure*}
\centering
\includegraphics[width=18.cm,clip]{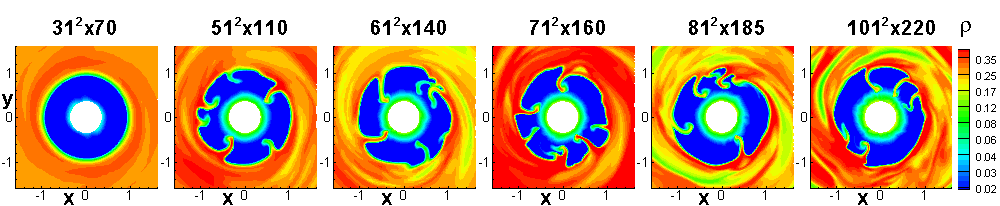}
\caption{ A comparison of simulation results for the model
$\mu0.5c1.5{\Theta5}\alpha0.02$ (at time $t=18$) using different
grid resolutions. Simulations show that accretion is stable in the
case of the coarsest grid, with $31^2\times70$ grids in one of the
six sides of the cubed sphere (\textit{left panel})   and is
unstable in the cases of the finer grids, $51^2\times110$,
$61^2\times140$, $71^2\times160$, $81^2\times185$, and
$101^2\times220$. The color background shows the density
distribution in the equatorial plane. } \label{grids-6}
\end{figure*}

\subsection{Dependence of instability on $\alpha$}
\label{sec:dependence-alpha}

In another set of experiments, we take the same model,
$\mu1c1.5\Theta5\alpha0.02$, and vary parameter $\alpha$ in the
range of $\alpha=0.02-0.1$. We observed that this initially stable
case becomes increasingly more unstable with increasing $\alpha$.
We are interested in knowing why this transition happens and why
the instability becomes stronger at higher values of $\alpha$.

Fig. \ref{sigav-alpha-r_cor} (right panel) shows results of our
simulations. One can see that at larger $\alpha$, the regime
switches from stable to unstable. The transition (where
$\gamma_{B\Sigma}^2 = \gamma_\Omega^2$) occurs at $\alpha\approx
0.037$. At larger values of $\alpha$, the disc-magnetosphere
boundary becomes more unstable due to both the decreasing
effective gravity $g_{\rm eff}$ and the increasing compression
factor $K_{\rm B\Sigma}$. The shear $\gamma_\Omega^2$ in the disc
is relatively small and varies slowly compared with the
$\gamma_{B\Sigma}^2$ term, thus having a smaller influence on the
transition.

It is interesting to note that the compression factor $K_{\rm
B\Sigma}$ increases with $\alpha$. This is probably due to the
fact that the radial velocity of matter in the disc increases with
$\alpha$: $v_r=\alpha c_s^2/v_K$ (where $c_s$ and $v_K$ are the
local sound speed and Keplerian velocity, respectively), and
higher gradients of surface density (per unit of magnetic flux)
are expected. On the other hand, the overall accretion rate
increases with $\alpha$: $\dot M\sim v_r\sim \alpha$, which leads
to a decrease in the magnetospheric radius: $r_m\sim 1/{\dot
M}^{2/7}$, and $g_{\rm eff}$ becomes more negative. The transition
between the stable and unstable regimes (where
$\gamma_{B\Sigma}^2=\gamma_\Omega^2$) occurs at $\alpha\approx
0.037$, which is near the point $\alpha=0.028$, where the
effective gravity $g_{\rm eff}$ changes the sign. Here, as in the
case of varying $r_{\rm cor}$, effective gravity plays an
important role in the transition between stable and unstable
regimes.

This set of simulations is similar to the simulations performed in
our earlier studies, where we investigated the transition between
stable and unstable regimes by changing parameter $\alpha$ to vary
the accretion rate $\dot M$, which we took to be the main factor
in determining the mode of accretion (e.g.,
\citealt{KulkarniRomanova2008,RomanovaEtAl2008}). In this paper,
our approaches were aimed at a deeper understanding of the physics
of transition between the stable and unstable regimes, and its
dependence on different parameters.

In realistic discs, the angular momentum transport is probably
provided by the magnetic turbulence, supported by the
magneto-rotational instability (\citealt{BalbusHawley1991}).  The
effective $\alpha-$parameter is determined by the ratio of
magnetic stress to matter pressure. The $\alpha-$ parameter may be
relatively high in the inner parts of the disc, where the magnetic
field of the turbulent cells is amplified by the rapid Keplerian
rotation of inner disc matter (e.g.,
\citealt{Hawley2000,Armitage2002}). In addition, part of the
stellar magnetic flux penetrates into the disc and increases the
magnetic stress in the inner disc, so that the $\alpha-$parameter
can be as high as $\alpha=0.3-1$
\citep{RomanovaEtAl2011,RomanovaEtAl2012,LiiEtAl2014}. This is why
we also performed test simulation runs at a relatively high
parameter of viscosity, $\alpha=0.1$. These simulations have shown
that instability is stronger when $\alpha=0.1$ than in the cases
of $\alpha=0.02$, and that the boundary between stable and
unstable regimes is expected to be higher, closer to the
$r_m/r_{\rm cor}\approx 1$ line, though additional research is
required to find this line.

\section{Dependence of instability on grid resolution}

\begin{figure}
\centering
\includegraphics[width=8.5cm,clip]{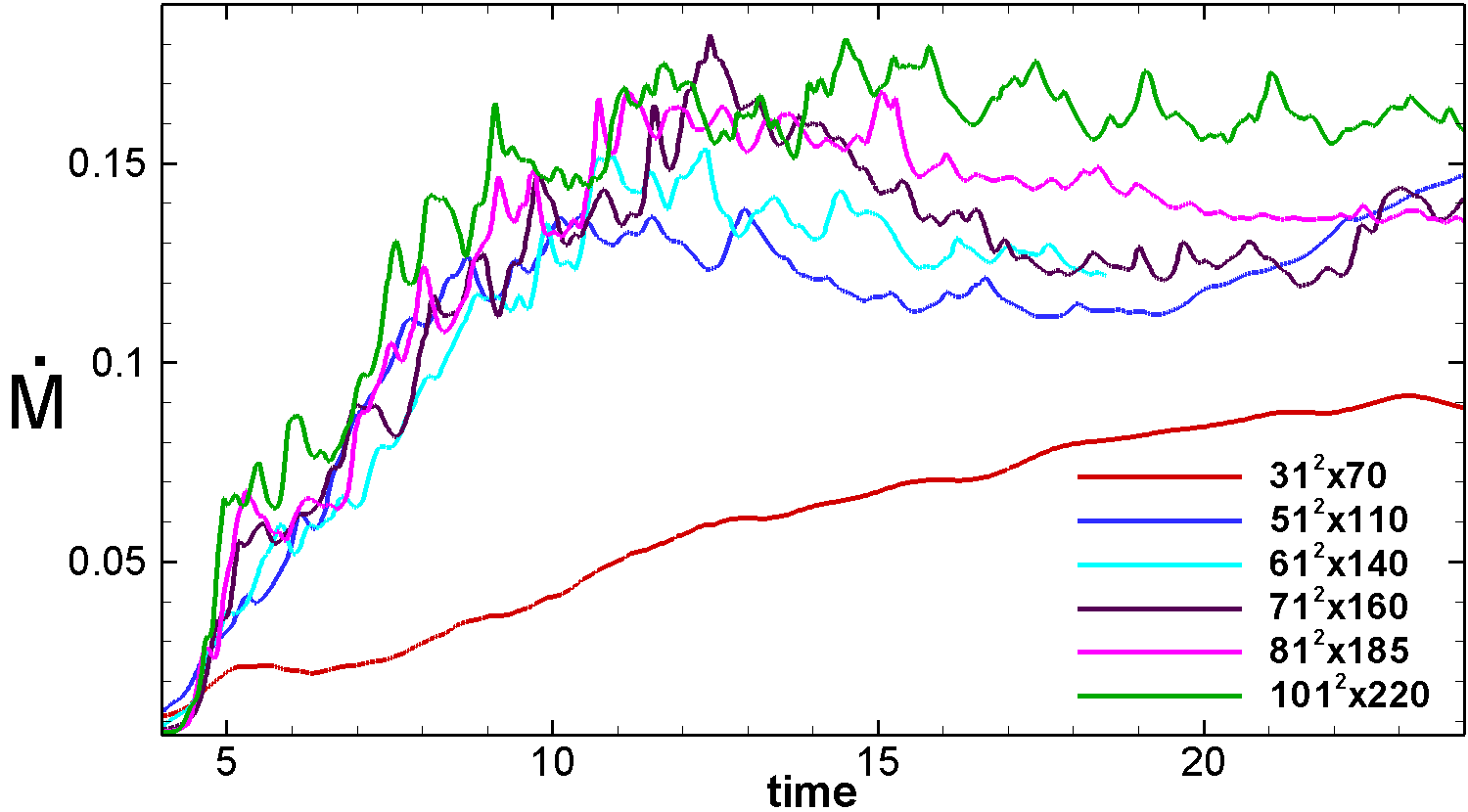}
\caption{Matter fluxes $\dot M$ onto the star as a function of
time for model $\mu0.5c1.5\Theta5\alpha0.02$, obtained at 6
different grid resolutions. }
\label{grids-pm}       
\end{figure}

To investigate the dependence of instability on grid resolution,
we chose a model with such parameters as make it close to the
boundary between stable and unstable regimes of accretion:
$\mu0.5c1.5\Theta5\alpha0.02$. First, we ran simulations of this
model using two grid resolutions, $31^2\times70$ and
$61^2\times140$. We noticed that the $31^2\times70$ case is
stable, while the $61^2\times140$ case is unstable. We then
performed a set of simulations of this model at several other grid
resolutions: $51^2\times110$, $71^2\times160$, $81^2\times185$,
and $101^2\times220$. We observed that instability is present at
all finer grid resolutions, including $51^2\times110$ and finer
(see Fig. \ref{grids-6}). We compared matter fluxes at all grid
resolutions, and found that while matter flux values are
substantially different when comparing the $31^2\times70$ and the
$61^2\times140$ cases, there is smaller
 difference in the matter flux values of all the finer
grid resolutions: $51^2\times110$, $61^2\times140$,
$71^2\times160$, $81^2\times185$, and $101^2\times220$ (see Fig.
\ref{grids-pm}). These comparisons show that at a coarse grid
resolution, numerical viscosity suppresses the development of the
RT instability, while at sufficiently fine grid resolutions, the
results do not depend much on grid resolution. Numerical viscosity
probably plays the same role as regular viscosity in suppressing
the smallest-scale perturbation modes, as discussed by
\citet{Chandrasekhar1961}.

\section{Dependence of instability on $\Theta$}
\label{sec:dependence-theta}

The above simulations were performed for a small tilt of the
dipole magnetosphere, $\Theta=5^\circ$. It is important to know
whether accretion through RT instability is present in the cases
of larger tilts. To answer this question, we performed a series of
simulation runs for a larger tilt, $\Theta=20^\circ$, with the
same goal in mind: to find the boundary between stable and
unstable regimes of accretion. Fig.~\ref{boundary-t5-t20} (right
panel) shows the results of simulations. We observed that
accretion becomes unstable in many instances, and the boundary
between stable and unstable regimes corresponds to the ratio
$r_m/r_{\rm cor} \approx 0.67$ (fastness parameter
$\omega_s\approx 0.54$). This line is very close to the boundary
line obtained for $\Theta=5^\circ$ (see dash-dot-dotted line right
above the bold line in the right panel of Fig.
\ref{boundary-t5-t20}). This result is in agreement with our
hypothesis that $g_{\rm eff}$ is the main factor in determining
the strength of the instability.

To better understand the unstable accretion onto stars with
different tilts of the dipole field, we performed test simulation
runs at different tilts of the dipole: $\Theta=10^\circ, 15^\circ,
20^\circ, 30^\circ, 40^\circ$ and $60^\circ$.
 We focused on the ordered unstable regime and
therefore used relatively small magnetospheres, $\mu=0.3-0.5$ and
large values of $r_{\rm cor}$. One of the questions was whether
this type of unstable accretion will take place in the cases of
larger tilts of the dipole. First, we performed simulation runs
for relatively small tilts, $\Theta=10^\circ, 15^\circ, 20^\circ,
30^\circ$.  Fig. \ref{theta-diff-6} shows $xy-$slices for two of
these angles, $\Theta=10^\circ$ and $20^\circ$. The xy-slices of
density distribution show ordered unstable accretion in one or two
tongues in the case of the smallest tilt, $\Theta=10^\circ$ (top
row of xy-slices). However, at a larger tilt, $\Theta=20^\circ$,
accretion becomes more chaotic (bottom row of xy-slices).

We also performed frequency analyses of the light-curves, which
were calculated from the moving hot spots on the stellar surface,
as seen at an inclination angle of $i=45^\circ$. The light-curve
for the $\Theta=10^\circ$ case (see second row in Fig.
\ref{theta-diff-6}) shows that the main source of variability is
the rotation of unstable ordered spots. However, modulation by
stellar rotation is also observed. The Fourier spectrum shows a
peak with a period of $P_{\rm inst}\approx 2$, associated with the
instability, and a peak associated with the period of the star,
$P_\star\approx 11.2$. The wavelet spectrum mainly shows the
frequency $f_{\rm inst}\approx 0.5$, associated with the
instability.

In the case of $\Theta=20^\circ$  the amplitude of the
oscillations associated with instability is smaller, and the
light-curve is mainly determined by the rotation of the star,
while the instabilities provide a high-frequency modulation of the
main light-curve. The presence of instability is weak in the
Fourier spectrum of the $\Theta=20^\circ$ case, but clearly
visible in the wavelet spectrum (see bottom row of Fig.
\ref{theta-diff-6}).

In the cases of even larger tilts of the dipole,
$\Theta=30^\circ$, $\Theta=40^\circ$ and $60^\circ$, the
light-curve is still modulated by the unstable component of
accretion, but the amplitude of the modulation is smaller than in
the case of $20^\circ$.

\section{Comparison between relativistic and non-relativistic
cases}

The results of current simulations are relevant for
non-relativistic stars. However, in case of neutron stars, we
should take into account the relativistic gravitational potential,
which provides a deeper gravitational well and may enhance
instability, in particular when the unstable tongues move closer
to the star. A relativistic potential has been used in most of our
prior simulations
\citep{KulkarniRomanova2008,KulkarniRomanova2009,RomanovaEtAl2008,BachettiEtAl2010}.

In this paper, we consider non-relativistic stars to be sure that
the instability mechanism is present even in the cases of
shallower, non-relativistic potential wells. However, in
application to neutron stars it is important to take into account
the relativistic effects and to compare the differences between
instabilities in non-relativistic and relativistic cases. It would
be too time-consuming to recalculate all the cases for
relativistic stars. Instead, we chose one representative model,
$\mu1c3\Theta5\alpha0.02$, and recalculated it taking into account
the relativistic corrections. To model the relativistic effects we
chose the Paczy\'nski-Wiita (PW) pseudo-relativistic potential
\citep{PaczynskiWiita1980}, $\Phi(r) = GM_\star/(r-r_g)$, where
$r_g=2GM\star/c^2$ is the gravitational (Schwarzschild) radius.
For a typical neutron star with mass $M_\star=1.4 M_\odot$ and
radius $R_\star=10$ km, the ratio $r_g/R_\star=0.145$.

\begin{figure}
\centering
\includegraphics[width=8.5cm]{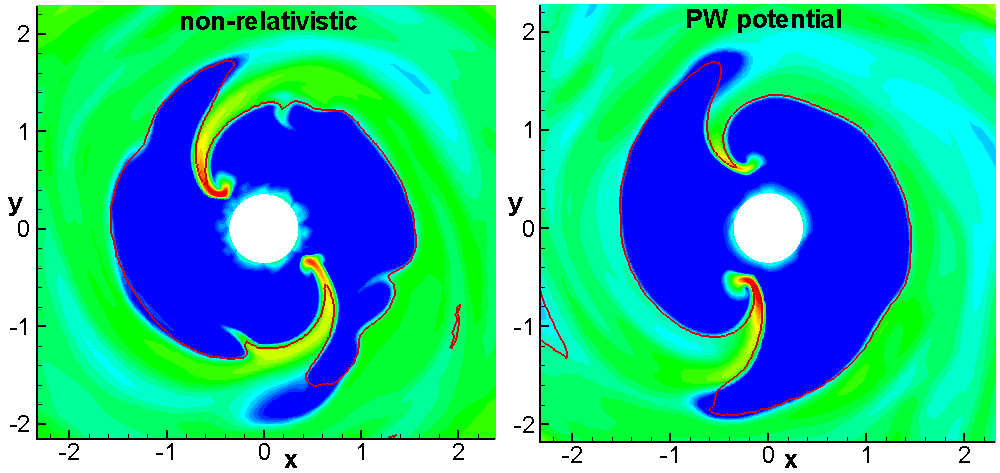}
\includegraphics[width=7cm]{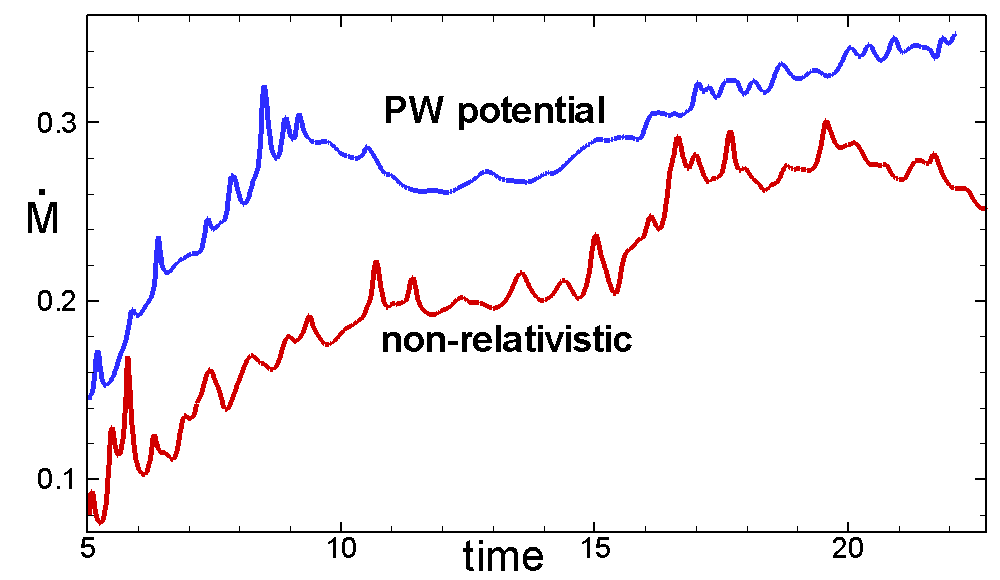}
\caption{\textit{Top panels:} density distribution of matter in
the equatorial (xy) slice in the model with $\mu=1$, $r_{\rm
cor}=3$ (model $\mu1c3\Theta5\alpha0.02$) at time $t=20$ in
non-relativistic (left) and relativistic (right) cases.
\textit{Bottom panel:} Matter fluxes to the star in
non-relativistic and relativistic cases.} \label{relat-nonrelat}
\end{figure}

We found the results to be similar: instability of comparable
strength has developed, with two ordered tongues forming and
rotating with the frequency of the inner disc.
Fig.\ref{relat-nonrelat} shows xy-slices of these two models for
the same moment in time. The matter flux onto the star (bottom
panel) is about 20\% higher in the relativistic case. However,
this does not affect the instability pattern qualitatively. We
should note that the PW potential gives a steeper gravitational
well than the fully relativistic approach, and therefore the
relativistic effects will be even weaker in a more realistic
relativistic case.

We conclude that the main effect of enhanced gravity is enhanced
accretion rate to the star, $\dot M$. However, the difference in
accretion rate between the non-relativistic and the PW
pseudo-relativistic cases is $\sim 20\%$, which leads to a factor
of only $1.2^{1/7}\approx 1.03$ difference in the magnetospheric
radius $r_m$. This difference is small, which is why relativistic
effects do not enhance instability significantly.

\section{Application to different magnetized stars} \label{sec:observations}
The results of our simulations can be applied to different types
of magnetized stars.

\subsection{Application to CTTSs}

Classical T Tauri stars stars show variability on different
time-scales, ranging from hours to months \citep{HerbstEtAl1994}.
In most cases, the causes of variability have not been understood.

Recent observations of the light-curves from young stars in the
NCG 2264 cluster, obtained with the \textit{CoRoT} telescope, show
that many stars have irregular variability
(\citealt{AlencarEtAl2010}). One of the important time-scales
corresponds to that expected in the unstable regime of accretion,
that is, a few peaks per rotational period of the inner disc
\citep{StaufferEtAl2014,CodyEtAl2014}.

The ordered unstable regime may also be important during the
periods of accretion outbursts, observed in EXor and FU Ori-type
stars (e.g., \citealt{AudardEtAl2014}). In these stars, the
accretion rate is greatly enhanced, the magnetosphere is
compressed, and the magnetospheric radius may be much smaller than
the corotation radius. Therefore, conditions become favorable for
ordered unstable accretion in one or two ordered tongues.
Recently, a very short period (of $\sim$ 1 day) was discovered in
the protostar V1647 Ori during its two recent accretion outbursts
\citep{HamaguchiEtAl2012}. Hamaguchi et al. (2012) suggested that
the puzzling short-term X-ray variability of V1647 Ori was
probably due to the rotational modulation of the hot spots on the
stellar surface. However, as the authors noted, the $\sim$ 1 day
period corresponds to rotation at near-breakup speed of V1647 Ori.
We suggest that it is improbable that a star rotates so rapidly.
Instead, the rapidly-rotating regions of enhanced plasma found in
the empirical model of \citet{HamaguchiEtAl2012} may be connected
with the rotation of the ordered unstable tongues in the ordered
unstable regime of accretion, which is highly likely to occur
during accretion outbursts.

\subsection{Application to accreting white dwarfs}

A few types of accreting white dwarfs have strong magnetic fields.
In the Intermediate Polars (hereafter IPs), the magnetosphere
disrupts the disc at $r_m\gtrsim 10 R_\star$, which is larger than
the radii considered in this paper. Simulations of accretion to
large magnetospheres \citep{RomanovaEtAl2014} show that matter of
the inner disc only partially penetrates the magnetosphere, and
most of the matter flows onto the star in two funnel streams
(above the magnetosphere), forming two ordered hot spots on the
surface of the star. In these stars, the light-curves reflect the
period of stellar rotation, and are expected to be periodic. This
process may explain the periodic light-curves of IPs.

There is another class of accreting white dwarfs, called Dwarf
Novae (DNs), that do not show period, but rather quasi-periodic
oscillations are observed during accretion outbursts
\citep{Warner2004}. The highest-frequency oscillations (called
Dwarf Novae Oscillations, or DNOs) have typical time-scales of
tens of seconds. The nature of these oscillations has not yet been
understood. It is possible that DNs are weakly magnetized white
dwarfs whose magnetospheres are strongly compressed during
accretion outbursts, with $r_{\rm cor}>>r_m$. This condition is
favorable for ordered unstable accretion, where one or two ordered
tongues are the cause of quasi-periodic variability. Variation in
accretion rate leads to variation in the position of the inner
disc and frequency of the tongues, which may possibly explain the
nature of DNOs.

\subsection{Application to accreting neutron stars}

The model can also be applied to accreting X-ray millisecond
pulsars (AMXPs), where matter of the inner disk accretes onto a
weakly magnetized ($B\sim 10^8-10^9$G) neutron star during
accretion outbursts
(\citealt{Vanderklis2006,IbragimovPoutanen2009,PatrunoWatts2012}).
Observations show pulsations associated with the rotation of the
star, and also one or two peaks associated with the high-frequency
quasi-periodic oscillations. The frequencies of QPOs vary in time
from values that are lower than the frequency of the star to
values that are much higher than the frequency of the star.  The
origin of these QPOs is still not well understood. We suggest
that, in AMXPs, one of the QPO frequencies may be associated with
the rotation of unstable tongues in the unstable regime of
accretion which onset when the accretion rate increases and the
magnetospheric radius approaches the value $r_m\approx 0.7 r_{\rm
cor}$. The chaotic tongues rotate with the frequency of the inner
disk, reflecting the QPO frequency (see also
\citealt{BachettiEtAl2010}). At even higher accretion rates, the
disk moves even closer to the star and, at $r_m\lesssim 0.6 r_{\rm
cor}$, the ordered unstable regime dominates. The quality factor,
Q, increases due the increasing dominance of the ordered unstable
regime. This is consistent with the observations, which show that
the frequencies of QPOs increase with X-ray flux (which is usually
proportional to the accretion rate)  (e.g.,
\citealt{PapittoEtAl2007}), and the quality factor increases with
increasing QPO frequency (e.g.,
\citealt{Vanderklis2006,BarretEtAl2007}).

We expect that the transition from the stable to the unstable
regime should lead to a decrease of the pulsed fraction in the
observed radiation (e.g., \citealt{KulkarniRomanova2008}). The
pulsations are associated with stable (two funnel stream)
magnetospheric accretion, where two hot spots form and rotate with
the frequency of the star. In contrast, in the unstable regime,
most of the matter flows in irregular tongues, and the expected
fraction of pulsed radiation is smaller. Interestingly, a drop in
the pulse fraction has been recently observed in the AMXP SAX
J1808.4-3658 \citep{BultVanDerKlis2015}. The pulse fraction
decreased dramatically when the QPO frequency (associated with the
frequency of the inner disk) increased and became comparable to
the stellar frequency. The drop in the pulse fraction may be
connected with the transition from stable to unstable accretion.
However, these observations show that the transition occurs when
$r_m\approx r_{\rm cor}$. This is different from the $r_m=0.7
r_{\rm cor}$ boundary found in our simulations at a small
parameter of viscosity ($\alpha=0.02$). Simulations show that at
larger values of $\alpha$ (e.g., $\alpha=0.1$) the boundary
corresponds to larger values of $r_m/r_{\rm cor}$. Our model can
explain the transition of the pulse fraction at high
$\alpha-$parameters of viscosity.

\section{Conclusions}

A new set of simulations has been performed with the goal of
investigating the boundary between stable and unstable regimes of
accretion and the properties of unstable accretion. Simulations
were performed at twice as high a grid resolution as in our
earlier studies
(\citealt{KulkarniRomanova2008,KulkarniRomanova2009,RomanovaEtAl2008}).
A low viscosity parameter in the disc  ($\alpha=0.02$) was used in
most of the simulation runs. The main results of the new
investigations are the following:

\textbf{1.} We found that the boundary between stable and unstable
regimes of accretion depends almost entirely on the fastness
parameter $\omega_s$ (or the ratio between the magnetospheric
radius and the corotation radius, $r_m/r_{\rm cor}$). Accretion is
unstable if $\omega_s\lesssim 0.6$ ($r_m/r_{\rm cor})\lesssim
0.71$), and is stable otherwise. The main simulations were
performed at a small misalignment angle of the dipole,
$\Theta=5^\circ$, and low viscosity in the disc, $\alpha=0.02$.

\textbf{2.} Below the   $\omega_s\approx 0.6$  ($r_m/r_{\rm
cor}\approx 0.71$) line, accretion proceeds in the \textit{chaotic
unstable regime} through several unstable tongues. The light-curve
from the hot spots looks chaotic, and the spectral analysis shows
several frequencies associated with the chaotic tongues. One of
the frequencies often corresponds to the frequency of the inner
disc, because the set of short-lived tongues rotates with the
angular frequency of the inner disc. The frequency of the star is
usually present in the frequency spectra if the simulation runs
are sufficiently long.

\textbf{3.} A new type of unstable regime of accretion, the
\textit{ordered unstable regime}, was found in the cases of slowly
rotating stars, $\omega_s\lesssim 0.45$ ($r_m/r_{\rm cor}\lesssim
0.59$). In this regime, matter accretes in one or two ordered
unstable tongues that rotate with the frequency of the inner disc.
The ordered unstable regime has been observed in the cases of
relatively small magnetospheres, $r_m/R_\star\lesssim 4.2$.

\textbf{4.} In the cases of larger misalignment angles of the
dipole, accretion through instabilities is also present. The
boundary between stable and unstable regimes was found for the
case of $\Theta=20^\circ$. This boundary, $\omega_s=0.55$, is very
close to the one obtained in the case of a small tilt,
$\Theta=5^\circ$, because in both cases the instability is mainly
determined by the effective gravity. However, in cases of
increasingly larger tilts, more matter flows above the
magnetosphere in regular funnel streams and forms two hot spots
near the magnetic poles, determining the regular sinusoidal
light-curve that represents the period of the star. The matter
accreting though instabilities provides the high-frequency
modulation of the main light-curve. The amplitude of the
oscillations decreases with $\Theta$ because a strongly tilted
dipole breaks the unstable tongues.

\textbf{5.} At a \textit{higher viscosity} in the disc,
$\alpha=0.1$, chaotic instability becomes more irregular, and
variability on different time-scales is observed in the
light-curve. The frequency associated with the inner disc rotation
is seen in both the Fourier and the wavelet spectra, while the
frequency of the star has a much lower amplitude.

\textbf{6.}  Analysis \textit{of the causes} of  instability in
the borderline cases between stable and unstable regimes shows
that: (a) Increasing the \textit{corotation radius} $r_{\rm cor}$
(while fixing all other parameters) leads to the larger negative
values of effective gravity $g_{\rm eff}$ causing the transition
from the stable to the unstable regime of accretion. We believe
the sign change of $g_{\rm eff}$ to be the main factor in this
transition. (b) Increasing the \textit{viscosity parameter}
$\alpha$ leads to a higher compression factor $K_{\rm B\Sigma}$
which leads to stronger instability.

\textbf{7.} A grid convergence has been observed at high grid
resolutions. Simulations show that instabilities are only
suppressed at the coarsest grid ($31^2$ angular cells in each
block), while at all finer grids ($51^2$ and finer) accretion
through instabilities is present, and is similar in all cases.

\textbf{8.} A comparison of relativistic and non-relativistic
cases shows that instability is somewhat stronger in the
relativistic cases. However, the relativistic potential is not the
main factor in determining the mode of accretion.

\textbf{9.} All of the above results were obtained for
magnetospheres with $r_m/R_\star\lesssim 7$ and should not be
generalized to stars with large magnetospheres. Separate studies
show that in the cases of large magnetospheres, instabilities are
only present in the external parts of the magnetosphere, while
matter accretes to the star in two ordered funnel streams
\citep{RomanovaEtAl2014}.

\textbf{10.} The results can be applied to different types of
accreting magnetized stars. The main findings are the following:

(a) Accretion in ordered unstable tongues can be important during
accretion outbursts, when the inner disc moves inward, compressing
the magnetosphere, and the ratio $r_m/r_{\rm cor}$ is small. The
frequency of the rotating tongues corresponds to the frequency of
the inner disc and may be seen as a QPO feature in the frequency
spectra. The QPO frequency increases when the inner disc moves
inward. The quality factor also increases during the inward motion
of the disc.

(b) When a star is observed for a few periods of stellar rotation,
as in CTTSs, the frequency of the oscillations associated with the
ordered unstable tongues may be mistaken for the frequency of the
star.

(c) A star may alternate between stable and unstable regimes of
accretion, thus showing an intermittency in its pulsations.


\subsection*{Acknowledgments}
{Authors thank Alexander Koldoba for an earlier-developed `cubed
sphere' code and the referee for critical reading of the paper.
Resources supporting this work were provided by the NASA High-End
Computing (HEC) Program through the NASA Advanced Supercomputing
(NAS) Division at Ames Research Center and the NASA Center for
Computational Sciences (NCCS) at Goddard Space Flight Center. The
research was supported by NASA grant NNX14AP30G and NSF grant
AST-1211318.}


\begin{appendix}

\section{Magnetospheric Radius}
\label{appendix-rm}

It is important to know the relationship between the
magnetospheric radii obtained  from the balance of stresses,
$\beta_1=1$ (which have been obtained from our 3D MHD numerical
simulations and used in our paper), and the magnetospheric radii
derived from the theoretical formula, \ref{eq:alfven}, which is
widely used for estimating $r_m$ (in the cases where the values of
$\mu_\star$, $\dot M$ and $M_\star$ are known). In our
simulations, we can compare these two radii by (1) taking one of
them from the condition $\beta_1=1$, and (2) taking the second one
from Eq. \ref{eq:alfven}, using parameters $\mu_\star$, $M_\star$
and $\dot M$, whose values are known from our simulations.
Comparisons of these radii will help us derive the unknown
coefficient $k$, as well as check the overall dependencies in Eq.
\ref{eq:alfven}.

For these comparisons, we first convert the theoretically-derived
magnetospheric radius (Eq. \ref{eq:alfven}) to the dimensionless
form (see procedure of dimensionalization in Sec.
\ref{sec:main-dim-parameters} and resulting Eq.
\ref{eq:rA-dimensionless}) \footnote{Here, we suggest that the
mass of the star is fixed at $M_\star=M_0$ so that the
dimensionless mass $\widetilde M_\star=1$. We also consider only
the cases with $\alpha=0.02$.}. We re-write Eq.
\ref{eq:rA-dimensionless} in a more general form:

\begin{equation}
r_m=k ({\mu}^2/{\dot{M}})^n~, \label{eq:rA-dimensionless-1}
\end{equation}
and search for coefficients $k$ and powers $n$. All variables are
dimensionless, but we removed the tilde's for convenience. Note
that coefficient $k$ is exactly the same as in Eq.
\ref{eq:alfven}, and the power $n$ can be different from $n=2/7$
used in the theoretical formula.

\begin{figure*}
\centering
\includegraphics[width=6.3cm]{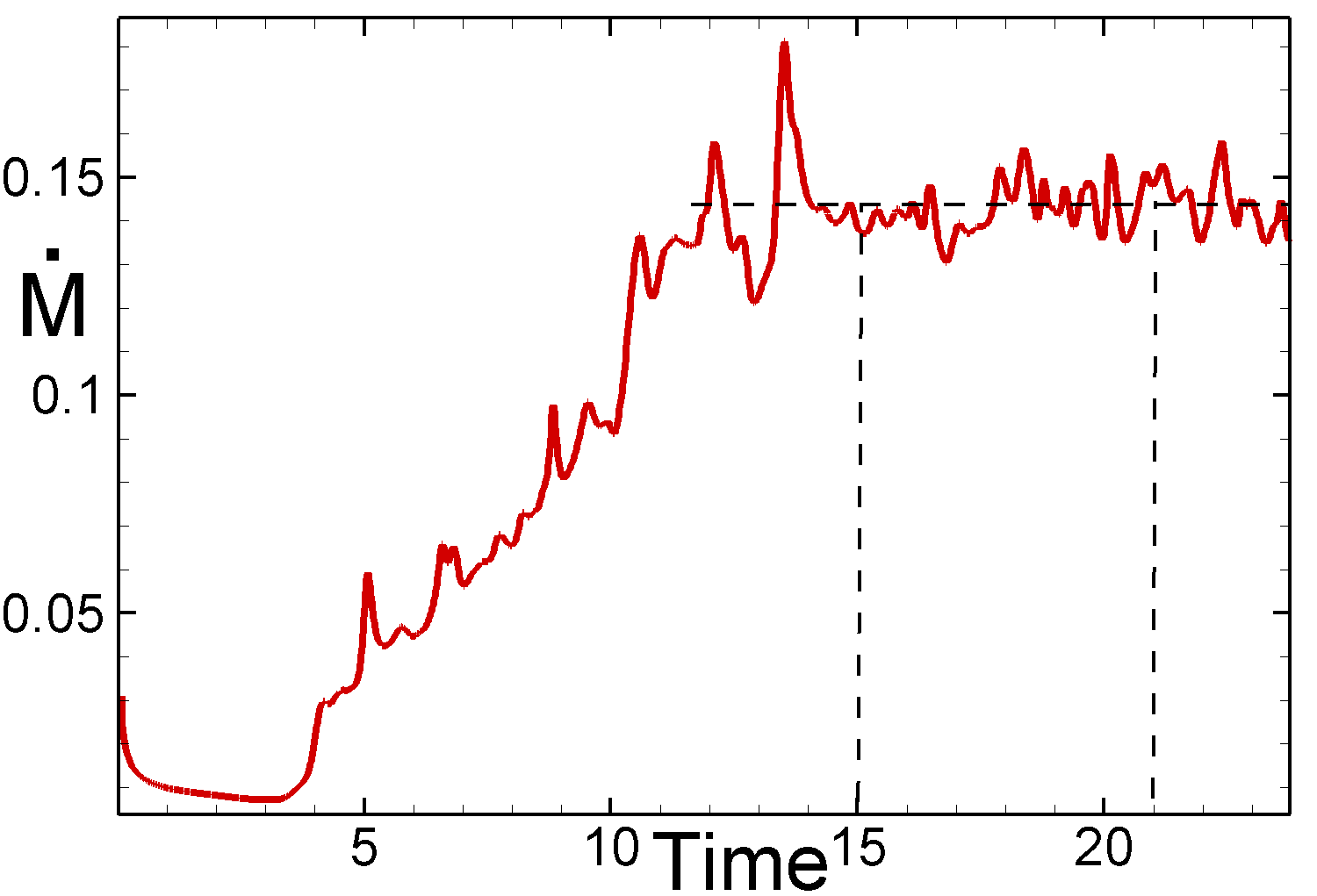}
\includegraphics[width=4.7cm]{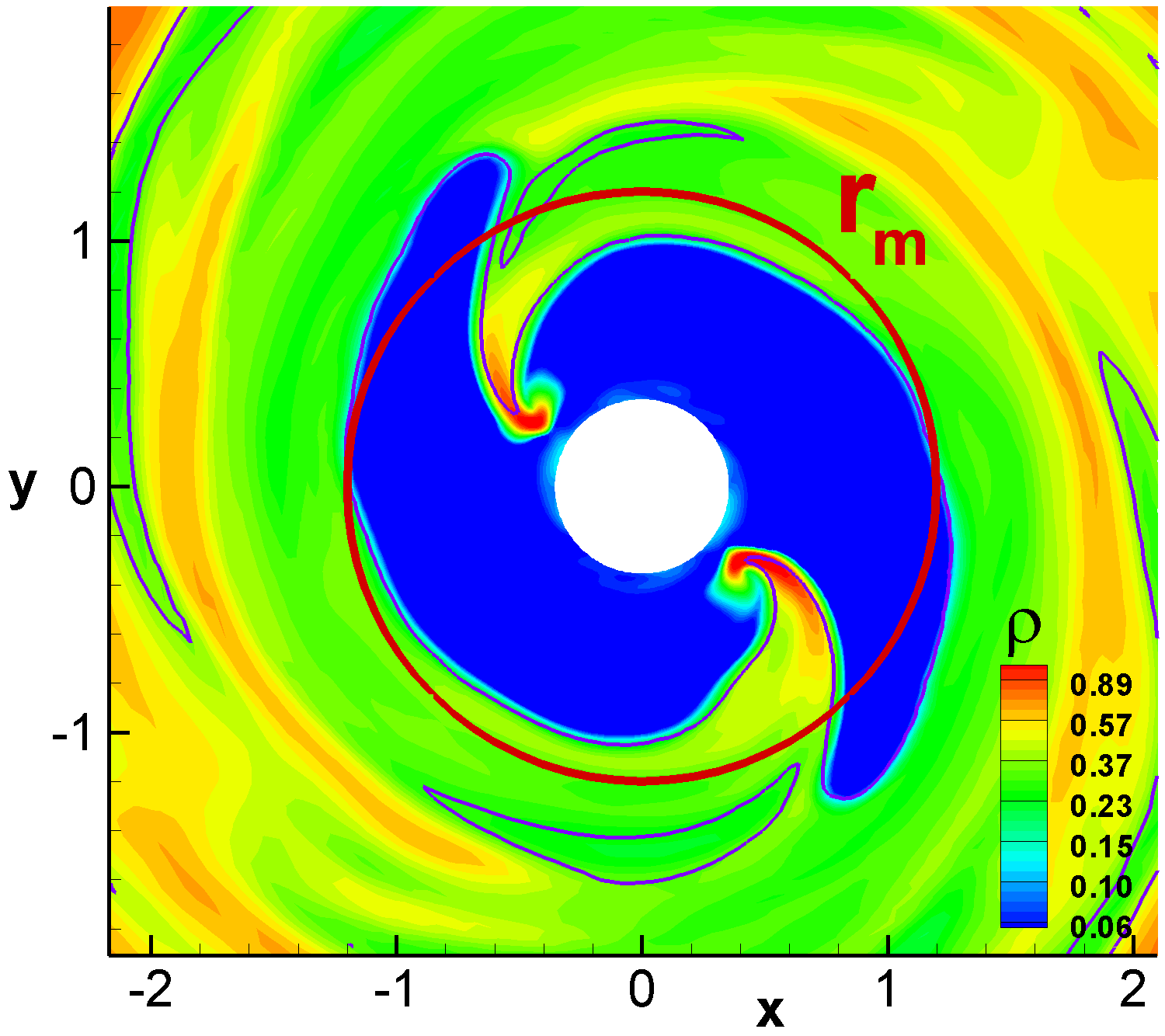}
\includegraphics[width=4.7cm]{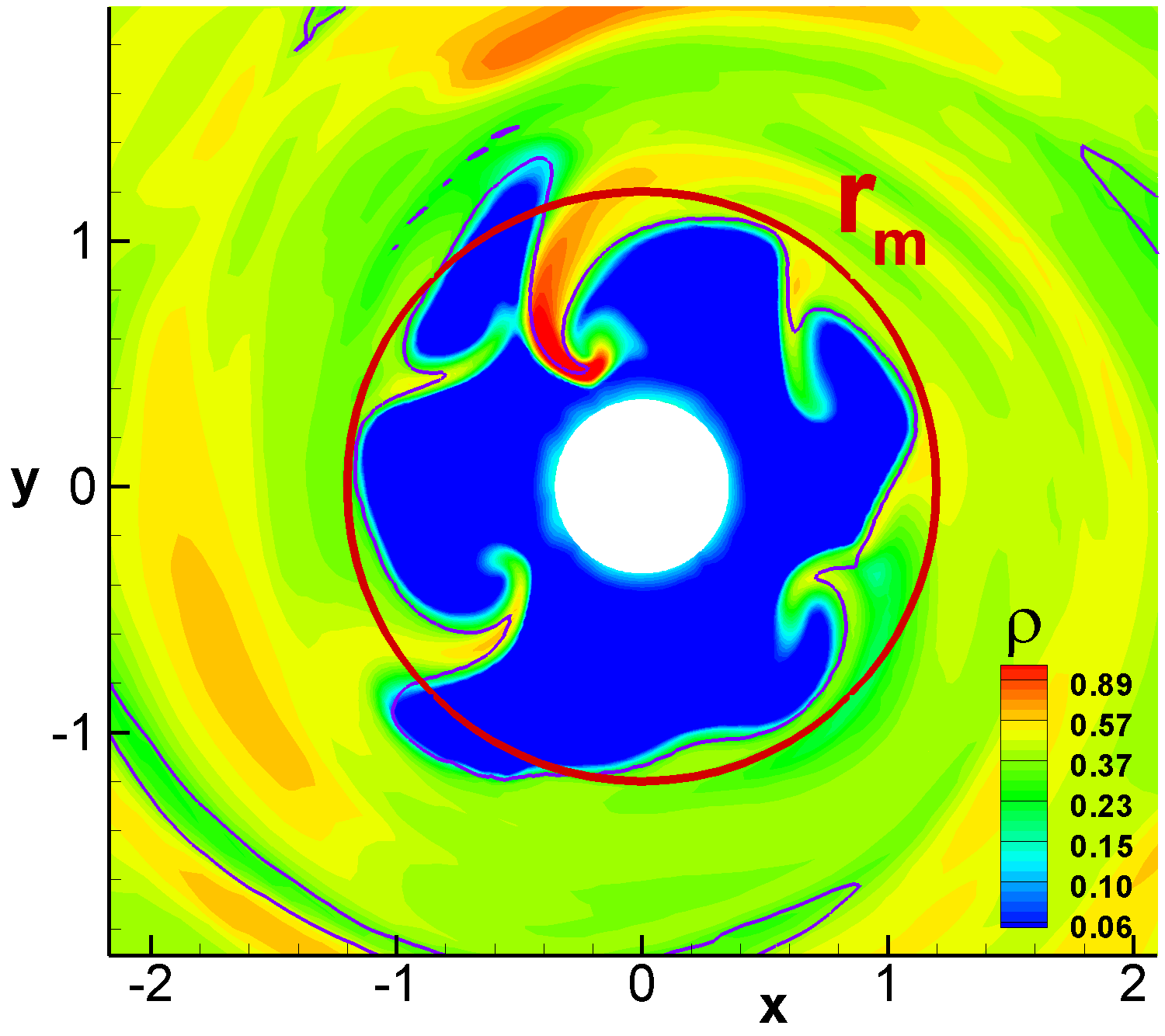}
\caption{\textit{Left panel:} Temporal variation of the
dimensionless matter flux $\dot M$ in one of models in the
unstable regime, $\mu0.5c2.5\Theta5\alpha0.02$. \textit{Middle and
right panels} show the equatorial slices of the density
distribution at time $t=15$ and $t=21$. The thin purple line
corresponds to condition $\beta_1=1$ (see Eq. \ref{eq:stress
balance}). Thick solid line shows the approximate position of the
magnetopsheric radius.} \label{app:rm-mdot}
\end{figure*}

Next, we take our models with different values of parameters $\mu$
and $r_{\rm cor}$, and find $r_m$ from the condition $\beta_1=1$.
We also measure the dimensionless matter flux $\dot M$ and
calculate the value $\mu^2/\dot M$. We then plot $r_m$ versus
$\mu^2/\dot M$. $\mu$ is a parameter of the model, and is fixed in
each model. However, the accretion rate $\dot M$ varies in time.
In some cases, the accretion rate levels off at a constant value,
in spite of the regime being strongly unstable, as shown in the
left panel of Fig. \ref{app:rm-mdot}. Alternatively, it may vary
strongly (by a factor of 2-3), as in the cases of the chaotic
unstable regime. This is why the main dispersion of points is
expected to arise from the variation of $\dot M$. We often took
several different points in time within the same simulation run to
see how the variation in accretion rate would affect the resulting
radius \footnote{$r_m$ varies in time, because stresses vary in
time. Note that finding $r_m$ does not require stationarity.}.

To measure the magnetospheric radius, we used the condition
$\beta_1=1$. However, in the unstable regime, the magnetospheric
radius often has a complex shape, and the $\beta_1=1$ line is not
truly circular. To obtain $r_m$ we approximate the inner boundary
using the procedure shown in Fig. \ref{app:rm-mdot}, where the
circle approximately reflects the inner boundary. We should note
that, in any particular simulation run, the value of $r_m$ does
not vary much, which may be the result of a weak (expected)
dependence on the accretion rate: $r_m\sim {\dot M}^{-2/7}$. For
example, variation of $\dot M$ by a factor of 3 leads to the
variation of $r_m$ by a factor of $1.3$.

\begin{figure*}
\centering
\includegraphics[width=7.cm]{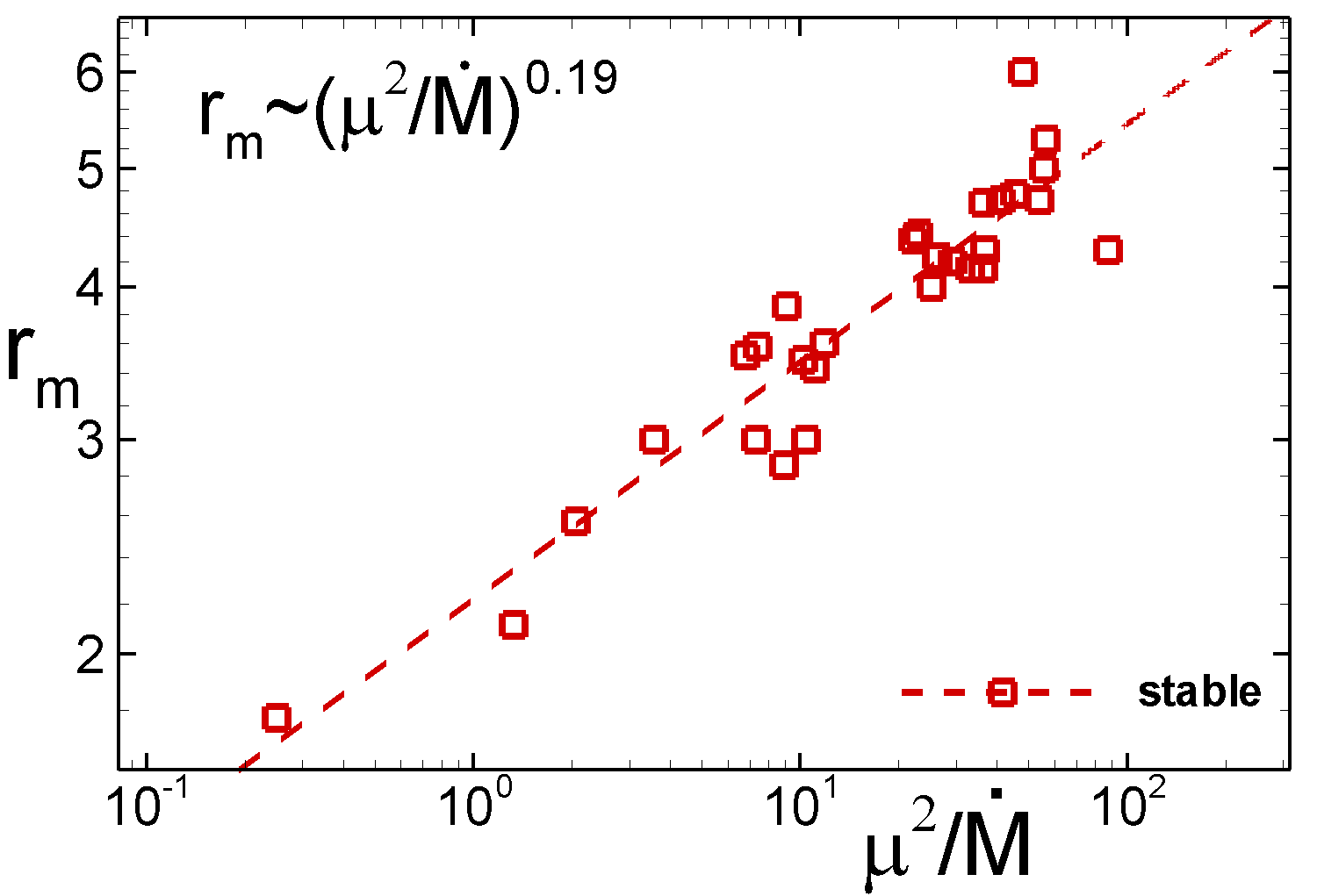}
\includegraphics[width=7.cm]{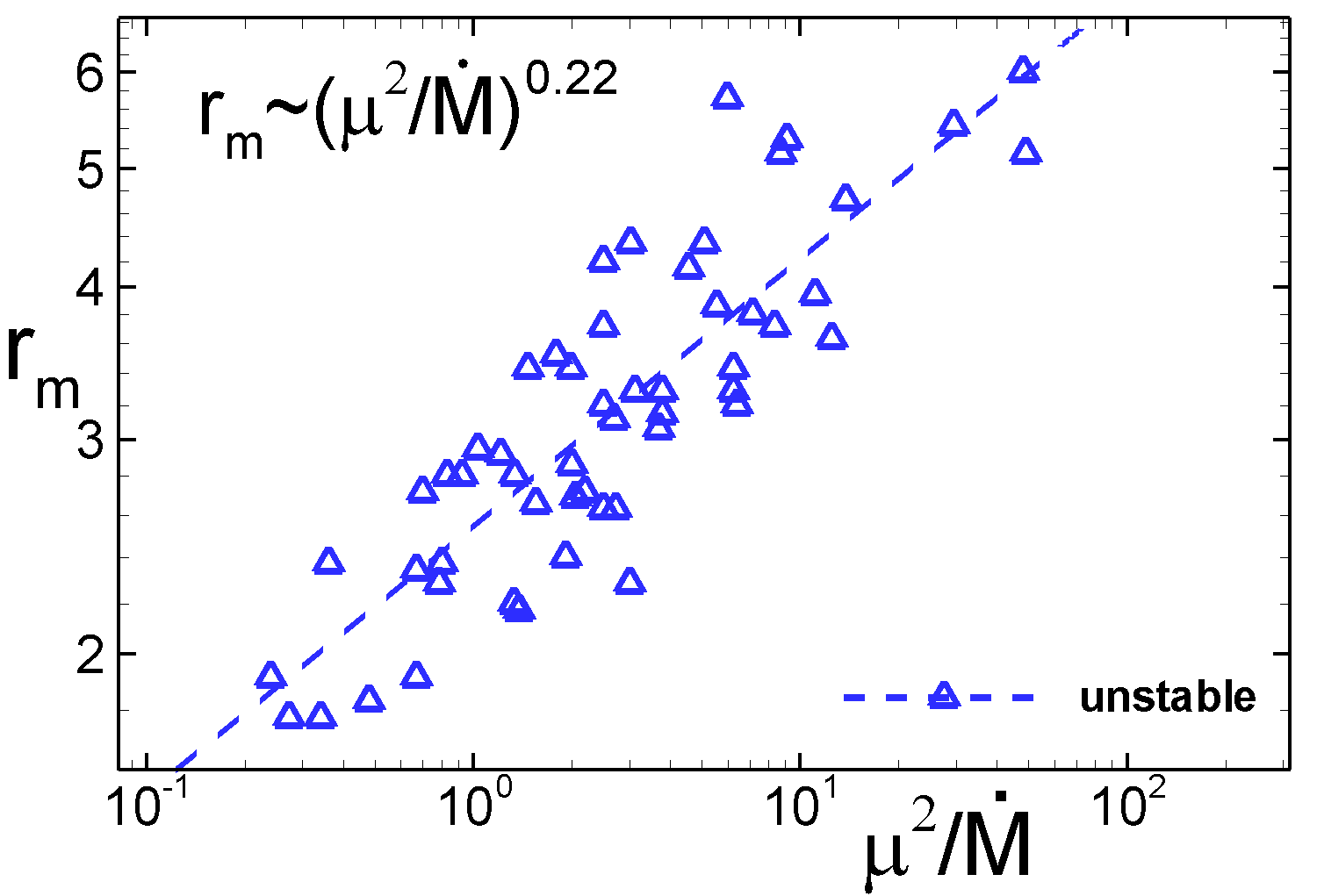}
\caption{\textit{Left panel:} The magnetospheric radii taken from
simulations in the stable regime, $r_m$ (measured in stellar
radius $R_\star$) versus the value $\mu^2/\dot M$. \textit{Right
panel:} same, but for the unstable regime. Dashed lines represent
the best power fit.} \label{app:rm-sim-theor}
\end{figure*}

Figure \ref{app:rm-sim-theor} shows the results of simulation runs
in the stable and unstable regimes (left and right panels,
respectively).
 One can see that in both cases the magnetospheric radius increases with
 parameter $\mu^2/\dot M$, and the best-fit curve is a power-law (see dashed
 lines), with the power of $n=0.22$ in the unstable regime and $n=0.19$
 in the stable regime. The plot also helped derive coefficient $k$,
 which is $k\approx 0.78$ in the stable regime and $k\approx 0.89$ in
 the unstable regime. Therefore, we expect the formula for the magnetospheric radius to be in the following form:

\begin{equation}{r_m}^{\rm stable} \approx  0.78 \big[\mu_\star^4/(\dot{M}^2
GM_\star)\big]^{0.095}  ~~, \label{eq:rm-stable}
\end{equation}
\begin{equation} {r_m}^{\rm unstable} \approx 0.89
\big[\mu_\star^4/(\dot{M}^2 GM_\star)\big]^{0.11}~~  .
\label{eq:rm-unstable}
\end{equation}
In both formulae, the power $n$ is smaller than the power
$1/7\approx 0.146$ in theoretical formula \ref{eq:alfven}. This
issue has been analyzed in detail by
\citet{KulkarniRomanova2013}\footnote{\citet{KulkarniRomanova2013}
performed similar comparisons of radii for 3D models in stable
regime, where the magnetic dipole is tilted by somewhat larger
angle, $\Theta=15^\circ$ and obtained the dependence $r_m\sim
({\mu}^2/{\dot{M}})^{0.2}$ . Our dependence for stable regime is
very close to one derived by these authors.}, who concluded that
the power $n$ is smaller than in the theoretical formula due to
the compression of the magnetosphere by the disc and the
non-dipole shape of the external regions of the magnetosphere.
This may also be the reason for why in the stable regime the power
$n$ is smaller than in the unstable regime: in the unstable
regime, the compression is expected to be smaller due to the
penetration of matter in unstable tongues.

We can estimate the difference between using the theoretical
formula \ref{eq:alfven} and formulae \ref{eq:rm-stable} or
\ref{eq:rm-unstable}. For comparisons, we take Eq.
\ref{eq:rA-dimensionless-1} in the form for stable regime,
$r_m=0.78({\mu}^2/{\dot{M}})^{0.19}$ and in the form corresponding
to Eq. \ref{eq:alfven}, $r_m=k (\mu^2/{\dot{M}})^{2/7}$ and find
$k=0.78 (\mu^2/{\dot{M}})^{-0.095}$. Then we note, that in our
model, $r_m/R_\star$ varies in the range of $2-6$ which
corresponds to
 $0.2\lesssim\mu^2/{\dot{M}}\lesssim 80$. For this range, we obtain $0.50\lesssim k\lesssim 0.90$.
 Similarly, for the unstable regime we find $k=0.89
(\mu^2/{\dot{M}})^{-0.026}$ and condition $0.80\lesssim k\lesssim
0.93$. One can see that in the case of small magnetospheres
$2\lesssim r_m/R_\star\lesssim 6$, theoretical formula
\ref{eq:alfven} can be used to describe the stable regime, if
$0.50\lesssim k\lesssim 0.90$, and the unstable regime, if
$0.80\lesssim k\lesssim 0.93$.

\section{Reference values in the Table}
\label{app:refval}

Table \ref{app:refval} shows sample values of physical parameters
for different types of stars. See Sec. \ref{sec:model} for a
detailed description.

\begin{table}
\begin{tabular}{llll}
\hline                & CTTSs              & White dwarfs   & Neutron stars      \\
\hline
$M_\star(M_\odot)$          & 0.8                & 1              & 1.4            \\
$R_\star$                 & $2R_\odot$         & 5000 km        & 10 km              \\
$B_\star$ (G)             & $10^3$             & $10^6$         & $3\e8$             \\
$R_0$ (cm)            & $4\e{11}$          & $1.4\e9$       & $2.9\e6$           \\
$v_0$ (cm s$^{-1}$)   & $1.6\e7$           & $3\e8$         & $8.1\e9$           \\
$\rho_0$ (g cm$^{-3}$) & $2.8\e{-11}$       & $7.9\e{-8}$      & $1.0\e{-5}$           \\
$\Sigma_0$ (g cm$^{-2}$) & $11.0$       & $112.5$          & $28.9$           \\
$\Omega_0$ (s$^{-1}$) & $4\e{-5}$          & 0.21           & $2.8\e3$           \\
$f_0$               & $0.55$ day$^{-1}$  & $3.2\e{-2}$ Hz & $4.5\e2$ Hz        \\
$P_0$                 & $1.8$ days         & 29 s           & 2.2 ms             \\
$\mu_\star$ (Gcm$^3$)       & $2.7\e{36}$          & $1.2\e{32}$      & $1.0\e{27}$     \\
$\dot M_0$ ($M_\odot$yr$^{-1}$) & $2.8\e{-7}$ & $1.9\e{-7}$ & $6.5\e{-8}$        \\
\hline \label{tab:refval}
\end{tabular}
\caption{Sample values of physical parameters for different types
of stars. See Sec. \ref{sec:model} for a detailed description.}
\label{tab:refval}
\end{table}

\end{appendix}

{}


\begin{thebibliography}{}

\bibitem[\protect\citeauthoryear{Alencar et al.}{2010}]{AlencarEtAl2010} Alencar, S.~H.~P. et~al., 2010, A\&A, 519, A88
\bibitem[\protect\citeauthoryear{Altamirano et al.}{2008}]{AltamiranoEtAl2008} Altamirano D., Casella P., Patruno A., Wijnands R., van der Klis M., 2008, ApJ, 674, L45
\bibitem[\protect\citeauthoryear{Armitage}{2002}]{Armitage2002} Armitage, P. 2002, ApJ, 330, 895
\bibitem[\protect\citeauthoryear{Arons \& Lea}{1976}]{AronsLea1976} Arons, J. \& Lea, S.M.  1976, ApJ, 207, 914
\bibitem[\protect\citeauthoryear{Audard et al.}{2014}]{AudardEtAl2014} Audard, M., {\'A}brah{\'a}m, P., Dunham, M. M., Green, J.D., Grosso, N. et al. 2014,  Accepted for publication as a review chapter in Protostars and Planets VI, University of Arizona Press (2014), eds. H. Beuther, R. Klessen, C. Dullemond, Th. Henning
\bibitem[\protect\citeauthoryear{Bachetti et al.}{2010}]{BachettiEtAl2010} Bachetti, M., Romanova, M. M., Kulkarni, A., Burderi, L., di Salvo, T. 2010, MNRAS, 403, 1193
\bibitem[\protect\citeauthoryear{Balbus \& Hawley}{1991}]{BalbusHawley1991} Balbus, S.A. \& Hawley, J. F. 1991, ApJ, 376, 214
\bibitem[\protect\citeauthoryear{Barret et al.}{2007}]{BarretEtAl2007} Barret, D., Olive, J,-F,, Miller, M. C. 2007, MNRAS, 376, 1139
\bibitem[\protect\citeauthoryear{Bessolaz et al.}{2008}]{BessolazEtAl2008} Bessolaz N., Zanni C., Ferreira J., Keppens R., Bouvier J. 2008, A\&A, 478, 155
\bibitem[\protect\citeauthoryear{Bouvier et al.}{2007}]{BouvierEtAl2007} Bouvier J.,  Alencar S. H. P., Harries T. J., Johns-Krull C. M., Romanova M. M., \textit{Protostars and Planets V}, Eds. Reipurth B., Jewitt D., Keil K. (University of Arizona Press, Tucson, 2007) 479
\bibitem[\protect\citeauthoryear{Bult \& van der Klis}{2015}]{BultVanDerKlis2015}  Bult, P., van der Klis, M. 2015, ApJ, 798, L29
\bibitem[\protect\citeauthoryear{Campbell}{1992}]{Campbell1992} Campbell, C.G. 1992, Geophys. Astrophys. Fluid Dynamics, 63, 179
\bibitem[\protect\citeauthoryear{Chandrasekhar}{1961}]{Chandrasekhar1961} Chandrasekhar, S., 1961, \textit{Hydrodynamic and Hydromagnetic Stability}. Clarendon, Oxford, p. 466
\bibitem[\protect\citeauthoryear{Cody et al.}{2014}]{CodyEtAl2014} Cody A.~M.,  Stauffer, J., Baglin, A., Micela, G., Rebull, L. M., Flaccomio, E., Morales-Calderón, M. et al., 2014, AJ, 147, 47 pp
\bibitem[\protect\citeauthoryear{Donati et al.}{2007}]{DonatiEtAl2007} Donati, J.-F., Jardine, M. M., Gregory, S. G., et al., 2007, MNRAS 380, 1297
\bibitem[\protect\citeauthoryear{Ghosh}{2007}]{Ghosh2007} Ghosh, P., 2007, Rotation and Accretion Powered Pulsars: World Scientific Series in Astronomy and Astrophysics -- Vol. 10. Edited by Pranab Ghosh. Published by World Scientific Publishing Co., Pte. Ltd., Singapore
\bibitem[\protect\citeauthoryear{Ghosh \& Lamb}{1978}]{GhoshLamb1978} Ghosh, P., Lamb, F. K., 1978, ApJ, 223, L83
\bibitem[\protect\citeauthoryear{Hamaguchi et al.}{2012}]{HamaguchiEtAl2012} Hamaguchi, K., Grosso, N., Kastner, J. H., Weintraub, D. A., Richmond, M. et al. 2012, ApJ, 754, 9pp
\bibitem[\protect\citeauthoryear{Hawley}{2000}]{Hawley2000} Hawley, J. F. 2000, ApJ, 528, 462
\bibitem[\protect\citeauthoryear{Hellier}{2001}]{Hellier2001} Hellier, C. 2001, Cataclysmic variable stars, (Springer, Berlin 2001)
\bibitem[\protect\citeauthoryear{Herbst et al.}{1994}]{HerbstEtAl1994} Herbst, W., Herbst, D. K., Grossman, E. J., Weinstein, D. 1994, AJ, 108, 1906
\bibitem[\protect\citeauthoryear{Ibragimov \& Poutanen}{2009}]{IbragimovPoutanen2009} Ibragimov, A. \& Poutanen, J., MNRAS, 400, 429
\bibitem[\protect\citeauthoryear{Illarionov \& Sunyaev}{1975}]{IllarionovSunyaev1975} Illarionov, A. F., \& Sunyaev, R. A. 1975, A\&A, 39, 185
\bibitem[\protect\citeauthoryear{Johns-Krull}{2007}]{Johns-Krull2007} Johns-Krull C. M., 2007, ApJ, 664, 975
\bibitem[\protect\citeauthoryear{Kaisig et al.}{1992}]{KaisigEtAl1992} Kaisig, M., Tajima, T., \& Lovelace, R. V. E. 1992, ApJ., 386, 83
\bibitem[\protect\citeauthoryear{Koldoba et al.}{2002}]{KoldobaEtAl2002} Koldoba, A. V.,  Romanova, M. M.,  Ustyugova, G. V., Lovelace, R. V. E. 2002, ApJ, 576,  L53
\bibitem[\protect\citeauthoryear{Koldoba et al.}{2008}]{KoldobaEtAl2008} Koldoba, A. V., Ustyugova, G. V., Romanova, M. M., Lovelace, R. V. E. 2008, MNRAS, 388, 357
\bibitem[\protect\citeauthoryear{Kulkarni \& Romanova}{2005}]{KulkarniRomanova2005} Kulkarni, A. K., \& Romanova, M. M., 2005, ApJ, 633, 349
\bibitem[\protect\citeauthoryear{Kulkarni \& Romanova}{2008}]{KulkarniRomanova2008} Kulkarni, A., \& Romanova, M.M. 2008, ApJ, 386, 673
\bibitem[\protect\citeauthoryear{Kulkarni \& Romanova}{2009}]{KulkarniRomanova2009} Kulkarni, A., \& Romanova, M.M. 2009, ApJ, 398, 1105
\bibitem[\protect\citeauthoryear{Kulkarni \& Romanova}{2013}]{KulkarniRomanova2013} Kulkarni, A., \& Romanova, M.M. 2013, MNRAS, 433, 3048
\bibitem[\protect\citeauthoryear{Kurosawa \& Romanova}{2013}]{KurosawaRomanova2013} Kurosawa R. and Romanova M. M., MNRAS 2013, 431, 2673
\bibitem[\protect\citeauthoryear{Lamb et al.}{1973}]{LambEtAl1973} Lamb, F.~K., Pethick, C.~J., Pines, D. 1973, ApJ, 184, 271
\bibitem[\protect\citeauthoryear{Lii et al.}{2014}]{LiiEtAl2014} Lii, P.S., Romanova, M.M., Ustyugova, G.V., Koldoba, A.V., Lovelace, R.V.E. 2014, MNRAS, 441, 86
\bibitem[\protect\citeauthoryear{Long et al.}{2005}]{LongEtAl2005} Long, M., Romanova, M.M., \& Lovelace, R.V.E. 2005, ApJ, 634,
1214
\bibitem[\protect\citeauthoryear{Long et al.}{2007}]{LongEtAl2007} Long, M., Romanova, M.M., \& Lovelace, R.V.E. 2007, MNRAS, 374, 436
\bibitem[\protect\citeauthoryear{Long et al.}{2008}]{LongEtAl2008} Long, M., Romanova, M.M., \& Lovelace, R.V.E. 2008, MNRAS, 386, 1274
\bibitem[\protect\citeauthoryear{Lovelace et al.}{1999}]{LovelaceEtAl1999} Lovelace, R.V.E., Romanova, M.M., Bisnovatyi-Kogan, G.S. 1999, ApJ, 514, 368
\bibitem[\protect\citeauthoryear{Lovelace et al.}{1995}]{LovelaceEtAl1995} Lovelace, R.V.E., Romanova, M.M., Bisnovatyi-Kogan, G.S. 1995, MNRAS, 275, 244
\bibitem[\protect\citeauthoryear{Lubow \& Spruit}{1995}]{LubowSpruit1995} Lubow, S.H., \& Spruit, H.C. 1995, ApJ, 445, 337
\bibitem[\protect\citeauthoryear{Orlando et al.}{2010}]{OrlandoEtAl2010} Orlando, S., Sacco, G. G., Argiroffi, C., Reale, F., Peres, G., Maggio, A. 2010, A\&A, 510,
12pp
\bibitem[\protect\citeauthoryear{Paczy\'nski \& Wiita}{1980}]{PaczynskiWiita1980} Paczy\'nski B. \& Wiita P. J. 1980, A\&A, 88, 23
\bibitem[\protect\citeauthoryear{Papitto et al.}{2007}]{PapittoEtAl2007} Papitto, A., di Salvo, T., Burderi, L., Menna, M. T., Lavagetto, G., Riggio, A. 2007, MNRAS, 375, 971
\bibitem[\protect\citeauthoryear{Patruno \& Watts}{2012}]{PatrunoWatts2012} Review to appear in ``Timing neutron stars: pulsations, oscillations and explosions", T. Belloni, M. Mendez, C.M. Zhang Eds., ASSL, Springer; arxiv1206.2727
\bibitem[\protect\citeauthoryear{Powell}{1999}]{PowellEtAl1999} Powell, K.G., Roe, P.L., Linde, T.J., Gombosi, T.I., \& De Zeeuw, D.L. 1999, J. Comp. Phys., 154, 284
\bibitem[\protect\citeauthoryear{Pringle \& Rees}{1972}]{PringleRees1972} Pringle, J.E., \& Rees, M.J. 1972, A\&A, 21, 1
\bibitem[\protect\citeauthoryear{Rast\"atter \& Schindler}{1999}]{RastatterSchindler1999} Rast\"atter, L. \& Schindler, K. 1999, ApJ, 524, 361
\bibitem[\protect\citeauthoryear{Romanova \& Kulkarni}{2009}]{RomanovaKulkarni2009} Romanova, M.M. \& Kulkarni, A.K. 2009, MNRAS, 398, 701
\bibitem[\protect\citeauthoryear{Romanova et al.}{2008}]{RomanovaEtAl2008} Romanova, M.M., Kulkarni, A.K., Lovelace, R.V.E. 2008, ApJ Letters, 273, L171
\bibitem[\protect\citeauthoryear{Romanova et al.}{2002}]{RomanovaEtAl2002} Romanova, M. M., Ustyugova, G. V., Koldoba, A. V., Lovelace, R.V.E., 2002, ApJ, 578, 420
\bibitem[\protect\citeauthoryear{Romanova et al.}{2003}]{RomanovaEtAl2003} Romanova, M. M., Ustyugova, G. V., Koldoba, A. V., Wick, J. V., Lovelace, R. V. E., 2003, ApJ, 595, 1009
\bibitem[\protect\citeauthoryear{Romanova et al.}{2004}]{RomanovaEtAl2004} Romanova, M. M., Ustyugova, G. V., Koldoba, A. V., Lovelace, R. V. E., 2004, ApJ, 610, 920
\bibitem[\protect\citeauthoryear{Romanova et al.}{2011}]{RomanovaEtAl2011} Romanova, M.M., Ustyugova, G.V., Koldoba, A.V., Lovelace, R.V.E. 2011, MNRAS, 416,
416
\bibitem[\protect\citeauthoryear{Romanova et al.}{2012}]{RomanovaEtAl2012} Romanova, M.M., Ustyugova, G.V., Koldoba, A.V., Lovelace, R.V.E. 2012, MNRAS, 421, 63
\bibitem[\protect\citeauthoryear{Romanova et al.}{2014}]{RomanovaEtAl2014} Romanova, M.M., Lovelave, R.V.E., Bachetti, M., Blinova, A.A., Koldoba, A.V., et al. 2014, Physics at the Magnetospheric Boundary, Geneva, Switzerland, Edited by E. Bozzo; P. Kretschmar; M. Audard; M. Falanga; C. Ferrigno; EPJ Web of Conferences, Volume 64
\bibitem[\protect\citeauthoryear{Romanova \& Owocki}{2015}]{RomanovaOwocki2015} Romanova, M.M., \& Owocki, S.P.  2015,
Space Science Reviews, 191, 339
\bibitem[\protect\citeauthoryear{Rucinski et al.}{2008}]{RucinskiEtAl2008} Rucinski S.~M. et al.,  2008, MNRAS, 391, 1913
\bibitem[\protect\citeauthoryear{Shakura \& Sunyaev}{1973}]{ShakuraSunyaev1973} Shakura, N.I., \& Sunyaev, R.A. 1973, A\&A, 24, 337
\bibitem[\protect\citeauthoryear{Stauffer et al.}{2014}]{StaufferEtAl2014} Stauffer, J., Cody, A.M., Baglin, A., Alencar, S., Rebull, L., Hillendbrand, L.A.  et al. 2014, AJ, 147, 34pp
\bibitem[\protect\citeauthoryear{Siwak et al.}{2011}]{SiwakEtAl2011} Siwak, M., Rucinski, S. M., Matthews, J. M., Pojmanski, G., Kuschnig, R. et al. 2011, MNRAS, 410, 2725
\bibitem[\protect\citeauthoryear{Spruit et al.}{1995}]{SpruitEtAl1995} Spruit H. C., Stehle R., Papaloizou J. C. B., MNRAS 1995, 275, 1223
\bibitem[\protect\citeauthoryear{Stauffer et al.}{2014}]{StaufferEtAl2014} Stauffer J. et al., 2014, AJ, 147, 34
\bibitem[\protect\citeauthoryear{Stone \& Gardiner}{2007a}]{StoneGardiner2007a} Stone J. M., Gardiner T. A., 2007a, Phys. Fluids, 19, 4104
\bibitem[\protect\citeauthoryear{Stone \& Gardiner}{2007b}]{StoneGardiner2007b} Stone J. M., Gardiner T. A., 2007b, ApJ, 671, 1726
\bibitem[\protect\citeauthoryear{van der Klis}{2006}]{Vanderklis2006} van der Klis M., \textit{Compact Stellar X-Ray Sources}, Eds. Lewin W. H. G. and van der Klis M. (Cambridge Univ. Press, Cambridge, 2006) 39
\bibitem[\protect\citeauthoryear{Wang \& Robertson}{1984}]{WangRobertson1984} Wang, Y.-M. \& Robertson, J.A. 1984, A\&A, 139, 93
\bibitem[\protect\citeauthoryear{Wang \& Robertson}{1985}]{WangRobertson1985} Wang, Y.-M. \& Robertson, J.A. 1985, ApJ, 299, 85
\bibitem[\protect\citeauthoryear{Warner et al.}{2004}]{Warner2004} Warner B., PASP 2004, 116, 115
\bibitem[\protect\citeauthoryear{Warner et al.}{1995}]{Warner1995} Warner B., 1995, Cataclysmic variable stars, (CUP, Cambridge 1995)

\end{thebibliography}
\end{document}